\documentclass[a4paper,fleqn]{cas-sc}

\usepackage[numbers]{natbib}

\usepackage{bm}
\usepackage{algorithmicx}
\usepackage{algorithm}
\usepackage{algpseudocode}

\usepackage{graphicx}
\usepackage{float}
\usepackage{subfig}
\usepackage{tabularx}
\def\tsc#1{\csdef{#1}{\textsc{\lowercase{#1}}\xspace}}
\tsc{WGM}
\tsc{QE}
\tsc{EP}
\tsc{PMS}
\tsc{BEC}
\tsc{DE}

\usepackage{microtype}  

\begin{document}
\let\WriteBookmarks\relax
\def\floatpagepagefraction{1}
\def\textpagefraction{.001}
\shorttitle{Design of a Turbo-based Deep Semantic Autoencoder for Marine Internet of Things}

\title [mode = title]{Design of a Turbo-based Deep Semantic Autoencoder for Marine Internet of Things}                      



\author[1]{Han}[prefix=Xiaoling]

\author[1]{Lin}[prefix=Bin,
                orcid=0000-0002-6125-9839]
\cormark[1]

\author[1]{Wu}[prefix=Nan]

\author[2]{Wang}[prefix=Ping]

\author[1]{Na}[prefix=Zhenyu]

\author[1]{Zhang}[prefix=Miyuan]

\affiliation[1]{organization={Department of Information Science and Technology, Dalian Maritime University},
                state={Dalian},
                country={China}}
\affiliation[2]{organization={Department of Electrical Engineering and Computer Science, York University,},
                city={Toronto},
                country={Canada}}
\cortext[cor1]{This manuscript has been accepted by Internet of Things, DOI: 10.1016/j.iot.2024.101393.}
\cortext[cor1]{Corresponding author \\ 
	\textit{E-mail address:} binlin@dlmu.edu.cn (B. Lin).}

\begin{abstract}
With the rapid growth of the global marine economy and flourishing maritime activities, the marine Internet of Things (IoT) is gaining unprecedented momentum. However, current marine equipment is deficient in data transmission efficiency and semantic comprehension. To address these issues, this paper proposes a novel End-to-End (E2E) coding scheme, namely the Turbo-based Deep Semantic Autoencoder (Turbo-DSA). The Turbo-DSA achieves joint source-channel coding at the semantic level through the E2E design of transmitter and receiver, while learning to adapt to environment changes. The semantic encoder and decoder are composed of transformer technology, which efficiently converts messages into semantic vectors. These vectors are dynamically adjusted during neural network training according to channel characteristics and background knowledge base. The Turbo structure further enhances the semantic vectors. Specifically, the channel encoder utilizes Turbo structure to separate semantic vectors, ensuring precise transmission of meaning, while the channel decoder employs Turbo iterative decoding to optimize the representation of semantic vectors. This deep integration of the transformer and Turbo structure is ensured by the design of the objective function, semantic extraction, and the entire training process. Compared with traditional Turbo coding techniques, the Turbo-DSA shows a faster convergence speed, thanks to its efficient processing of semantic vectors. Simulation results demonstrate that the Turbo-DSA surpasses existing benchmarks in key performance indicators, such as bilingual evaluation understudy scores and sentence similarity. This is particularly evident under low signal-to-noise ratio conditions, where it shows superior text semantic transmission efficiency and adaptability to variable marine channel environments.
\end{abstract}




\maketitle

\section{Introduction}

With the rapid development of the global marine economy and the increase in maritime activities, the marine Internet of Things (IoT) is becoming the enabler of all activities \cite{diyipian}. Traditional maritime communication systems, such as the Global Maritime Distress and Safety System (GMDSS), form the backbone of maritime communication, providing essential functions for ship communication, navigation, and emergency contact \cite{GMDSS1993, GMDSS2017}. However, communication issues in the marine IoT have been central to its development, necessitating significant enhancements in the reliability, efficiency, and stability of maritime communication systems to meet the application requirements of marine IoT \cite{zhihui6g}. The emergence of intelligent devices such as Unmanned Aerial Vehicles (UAVs), Unmanned Surface Vessels (USVs), and smart buoys introduces novel challenges, as traditional communication systems fall short in addressing the complex marine environments and substantial data transfer demands these devices entail. Therefore, there is an urgent need to improve the communication efficiency and reliability of maritime communication systems to guarantee safe navigation, environmental surveillance, and task accomplishment within the realm of maritime IoT.

In recent years, the integration of Artificial Intelligence (AI), particularly through the extensive application of Deep Learning (DL), has been a catalyst for significant advancements in the physical layer of communication systems. One of the main benefits of this integration is the ability to generalize to diverse channel environments, thereby providing a robust framework for data transmission and reception. This is achieved through the use of autoencoders, which learn the probability distribution of the channels without the need for explicit data understanding \cite{sadeghi2019physical, aoudia2019model, lu2020deep}. For instance, a Convolutional Neural Network (CNN)-based autoencoder communication system \cite{MAEACESS} was introduced, designed to adapt intelligently to various block lengths, accommodate diverse throughput needs, and perform effectively in both AWGN and Rician fading channels. Further, a novel orthogonal frequency-division multiplexing autoencoder based on CNN \cite{MAEIOT} was engineered, integrating channel estimation to thrive in the complex and volatile maritime environments, particularly in the context of marine IoT. Additionally, an End-to-End (E2E) CNN-based autoencoder \cite{CNN-AE} was devised to retain the characteristics of traditional systems while leveraging the advantage of CNN to jointly optimize various modules. Lastly, the CNN-based channel feedback autoencoder \cite{CNN-CF-AE} was developed, enhancing communication reliability through the incorporation of a feedback encoder and decoder.

The aforementioned communication systems primarily operate on a syntactic level, which limits their potential to achieve higher efficiency \cite{weaver}. Consequently, researchers are increasingly focusing on designing systems that understand the meaning of the messages rather than merely transmitting meaning-agnostic bits. As communication needs evolve, semantic communication is gaining significant attention \cite{tiaozhan}. This paradigm shift aims to transcend mere bit transmission by enabling systems to comprehend and process the true meaning encapsulated within messages \cite{ping2022intellicise, yuyichaoyuebite}. In the realm of marine IoT, the proliferation of UAVs, USVs, and other intelligent devices has amplified the demand for real-time, efficient, and reliable communication. These devices need to transmit and understand the meaning behind the data to make timely and accurate decisions. Therefore, the complexity and data-intensive nature of marine IoT require communication systems to not only be syntactically accurate but also capable of understanding and conveying the deeper meaning of information. This is essential for achieving efficient and intelligent interactions \cite{kaifangwenti}. Incorporating semantic communication into marine IoT can significantly elevate the intelligence level of maritime operations, enabling more effective navigation, environmental monitoring, and task execution \cite{yuyiwenti}.

The core idea of a semantic communication system is to integrate semantic coding modules, which enhance the system's ability to comprehend and convey the meaning of transmitted data. Researchers have proposed various DL-based solutions to achieve this. 
For instance, Xie \textit{et al.} proposed DeepSC, which leverages transformer technology to maximize system capacity and prioritize the recovery of sentence semantics over traditional bit or symbol error correction \cite{DeepSC}. L-DeepSC, developed by the same team, is a lightweight distributed semantic communication system that emphasizes low-complexity text transmission, improving semantic-level transmission efficiency and promoting data transfer from IoT devices to the cloud or edge \cite{L-DeepSC}. Peng \textit{et al.} introduced R-DeepSC, a robust DL semantic communication system that utilizes calibrated self-attention mechanisms and adversarial training to effectively address semantic noise, showing significant performance improvements over benchmarks focused solely on physical noise in text transmission, especially under various signal-to-noise ratio (SNR) conditions \cite{R-DeepSC}. Furthermore, Jiang \textit{et al.} proposed a knowledge graph-based semantic communication system that dynamically adjusts content transmission based on channel quality, substantially enhancing communication reliability under low SNR conditions \cite{zhishitupu}. Lastly, Hu \textit{et al.} introduced MR-DeepSC, a one-to-many semantic communication system that achieved superior performance in low SNR environments, surpassing other benchmarks \cite{MR_DeepSC}. But these schemes fail to protect the semantic information during transmission.

On the other hand, traditional error correction codes such as Turbo, Polar, and LDPC have been the cornerstone of reliable communication systems. Turbo codes, in particular, have garnered significant attention for their robustness in error correction, evidenced by their widespread application in satellite, mobile, and digital broadcasting communications. The integration of DL into these systems has shown remarkable potential. For example, Jiang \textit{et al.} designed a novel DL architecture named DEEPTURBO specifically for Turbo decoding, which offers excellent error rate performance and low training cost \cite{DEEPTURBO}. In another study, they introduced TurboAE, an E2E neural encoder and decoder that are jointly trained, approaching state-of-the-art performance under standard channel settings and demonstrating outstanding reliability under non-standard configurations \cite{jiang2019turbo}. They also proposed FTAE, a feedback autoencoder that integrates interleaving and iterative decoding with a CNN architecture, improving performance in terms of block length and block error feedback settings \cite{FTAE}. Further advancements include TurboAE-TI, which integrates TurboAE with a trainable interleaver design, showcasing benchmark performance advantages under various channel conditions \cite{TurboAE-TI}. He \textit{et al.} introduced TurboNet, which unfolds the original iterative structure into deep neural network decoding units, highlighting its powerful learning capability and resilience in diverse scenarios \cite{TurboDNN}. Lastly, Hebbar \textit{et al.} introduced TINYTURBO, a neural-enhanced Turbo code decoder that approaches the performance of the MAP algorithm and surpasses the maximum log-MAP baseline \cite{TinyTurbo}. However, all these DL-based turbo schemes are designed to minimize Bit Error Rate (BER). In other words, these schemes are unable to capture the meaning of the text through its structure.

Building on the previously discussed analysis, this paper is dedicated to developing an efficient maritime E2E text semantic transmission solution. More explicitly, we propose a Turbo-based deep semantic autoencoder (Turbo-DSA) for marine IoT. This scheme embeds Turbo coding structure using DL technology with the transformer to constitute the nucleus of the proposed Turbo-DSA. Our method transplant the error correction capability of Turbo codes to the needs of semantic information, focusing on extracting and reconstructing the semantic features of maritime text to provide high-quality maritime semantic communication services. The overall design of Turbo-DSA is formulated as an optimization problem, aiming to maximize E2E semantic transmission performance, rather than BER. The specific contributions of this paper are as follows.

\begin{figure}
	\centering 
	\vspace*{0pt} 
	\includegraphics[width=0.8\linewidth]{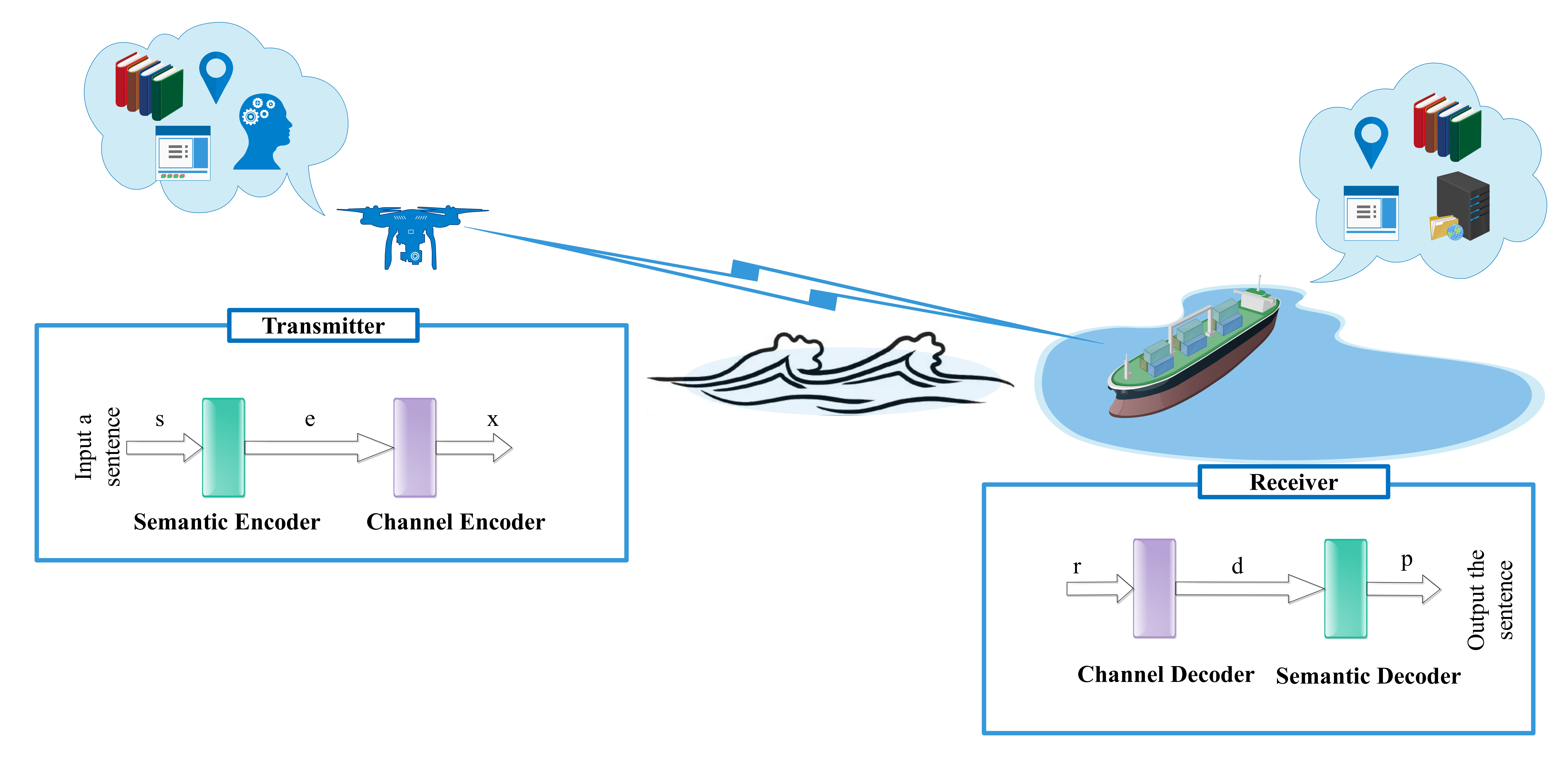}
	\caption{The system model of the proposed Turbo-DSA.} 
	\label{TurboDSA} 
\end{figure}

\begin{enumerate}

    \item A novel coding scheme, named Turbo-DSA, featuring semantic understanding of the transmitted content, is proposed. 
    Its joint source and channel coding specifically focuses on the semantic extraction, protection, transmission, and recovery of text messages. The transmitter comprises a semantic encoder and a channel encoder based on Turbo principles, while the receiver features an iterative channel decoder and a semantic decoder, ensuring efficient overall transmission of semantic information.
    
    \item The Turbo-DSA is benefiting from the interaction design of its components.
    The semantic encoder and decoder, composed of transformers, effectively transform messages into semantic vectors. Furthermore, it is capable of capturing the underlying semantic context of the background knowledge base and adapting to channel characteristics. The channel encoder employs the Turbo principle to distance these vectors in the semantic space, ensuring precise transmission even under channel turbulence. Meanwhile, the channel decoder utilizes Turbo iterative decoding to benefits from representation of semantic vectors.
    
    \item The novel efficient training and testing algorithms, tailored for the turbo-enhanced neural network structure, are proposed.
    The deep integration of the transformer with the Turbo framework significantly increases the training difficulty. By leveraging forward and backward propagation, the training process of the proposed Turbo-DSA achieves rapid convergence. Furthermore, the proposed testing algorithm rigorously evaluates system performance under various channel conditions.

    \item Extensive simulation results showcasing the remarkable abilities of the Turbo-DSA are provided.
    Numerous simulation results indicate that the proposed Turbo-DSA scheme exceeds benchmark models in terms of semantic extraction and recovery of maritime text across a wide range of SNRs, showcasing its generalization capabilities and robustness.
    
\end{enumerate}

The structure of the remainder of this paper is as follows. The proposed Turbo-DSA is introduced in Section \ref{Turbo-DSA}, including semantic encoder, channel encoder, channel decoder, semantic decoder and the loss function. Detailed training and testing algorithms are presented in Section \ref{Algorithm}. Simulation experiments for the proposed Turbo-DSA are demonstrated in Section \ref{Simulation}. Finally, we conclude this paper in Section \ref{Conclusion}.

\section{System Model}
\label{Turbo-DSA}

To achieve transmission of information in the changing marine environment, we develop a novel Turbo-DSA. As depicted in Fig. \ref{TurboDSA}, the model establishes a complete E2E communication system, tailored to the needs of maritime communication. The UAV, serving as the transmitter, is equipped with a semantic encoder and channel encoder, focusing on precisely capturing and encoding the semantic features of the messages. At the receiver, the USV is equipped with a channel decoder and a semantic decoder, which not only decode the received signals but also conduct an in-depth analysis of the underlying meanings of the signals. 

Turbo-DSA employs DL to transition information processing from the syntactic level to the semantic level. Concurrently, the transmitter and receiver collaborate to facilitate the integration of source coding and channel coding at the semantic level. This not only retains the core semantics of the original information but also enhances the semantic expressiveness of the information. Furthermore, the design of Turbo-DSA takes into account the dynamics of the marine environment, endowing the model with the ability to adapt to environmental changes. It continuously evolves its communication strategies through ongoing learning to optimize performance in the ever-changing marine environment. A key assumption of this work is that both the transmitter and receiver have access to the same background knowledge base, ensuring consistency in semantic understanding.

\begin{figure*}
   \centering 
   \vspace*{0pt} 
   \includegraphics[width=0.8\linewidth]{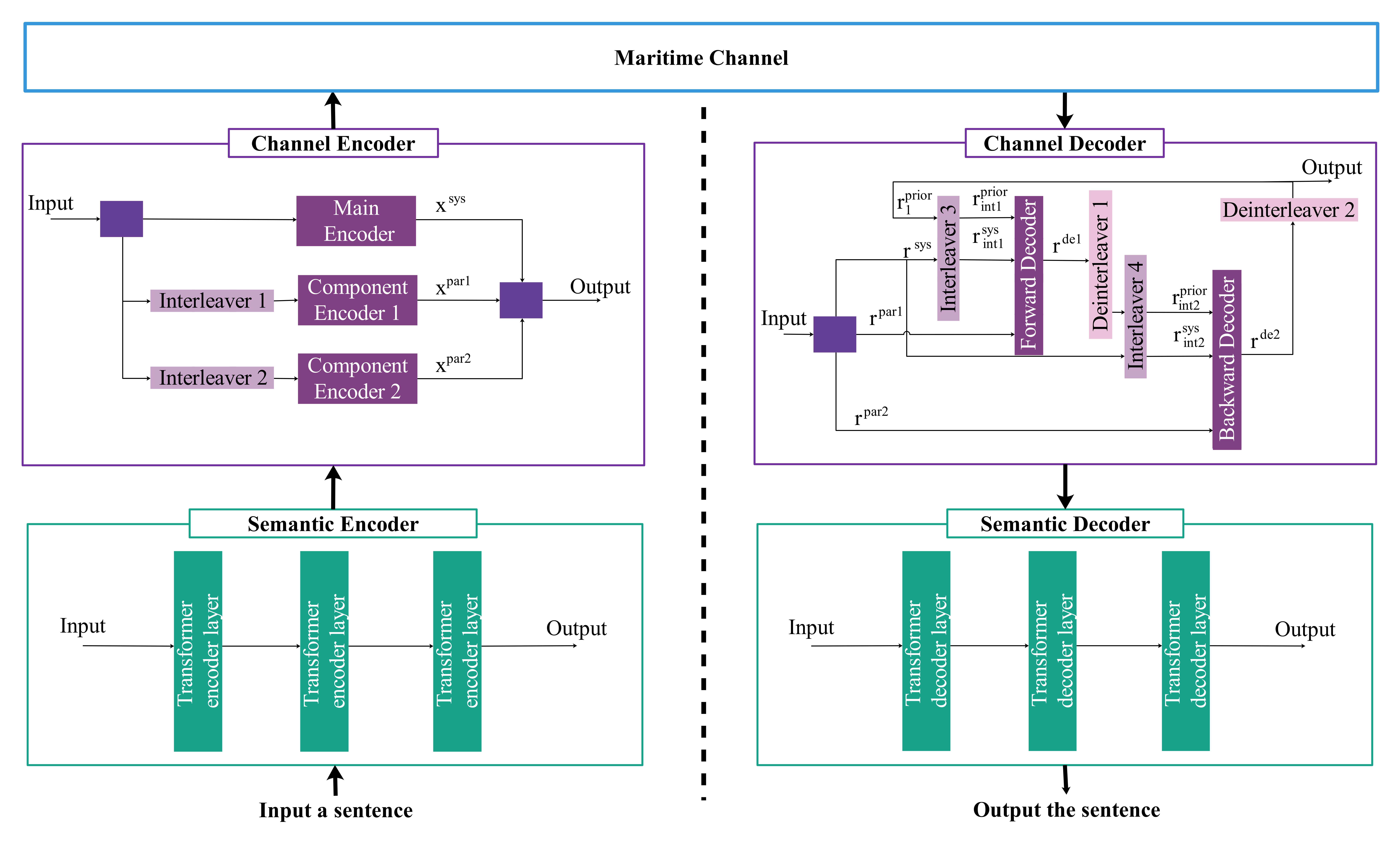}
   \caption{The signal transmission process of the proposed Turbo-DSA.} 
   \label{TurboSignal} 
\end{figure*}

The structure and signal transmission process of the proposed Turbo-DSA are depicted in Fig. \ref{TurboSignal}. Concurrently, Tab. \ref{structuralparameters} displays the structure of the proposed Turbo-DSA, where $ B $ represents the batch size, $ L $ represents the number of words in a sentence, $ D_1 $, $ D_2 $, $ D_3 $ represent the output dimensions of different network layers, $ D_4 $ represents the dimension of the prior information sequence, and $ D_5 $ represents the number of words in the corpus. Below, we will provide a detailed introduction to the structure and signal processing process of the proposed Turbo-DSA.

\begin{table}
    \centering
    \caption{The structural parameters of the proposed Turbo-DSA.}\label{structuralparameters}
    \begin{tabularx}{0.8\textwidth}{>{\centering\arraybackslash}p{0.2\textwidth} >{\centering\arraybackslash}p{0.2\textwidth} >{\centering\arraybackslash}p{0.4\textwidth}}
    \hline
        \textbf{Layer category} & \textbf{Layer name} & \textbf{Output dimension}  \\ \hline
        \multirow{4}{*}{Semantic encoder} & Embedding layer & [$B$, $L$, $D_1$]  \\ 
        ~ & Transformer encoder 1 & [$B$, $L$, $D_1$]  \\ 
        ~ & Transformer encoder 2 & [$B$, $L$, $D_1$]  \\ 
        ~ & Transformer encoder 2 & [$B$, $L$, $D_1$]  \\ 
        \hline
        \multirow{8}{*}{Turbo encoder} & Main encoder & [$B$, $L$, $D_2$]   \\ 
        ~ & Linear & [$B$, $L$, $D_2$]    \\ 
        ~ & Interleaver 1 & [$B$, $L$, $D_1$]  \\ 
        ~ & Component encoder 1 & [$B$, $L$, $D_2$]   \\ 
        ~ & Linear & [$B$, $L$, $D_2$]   \\ 
        ~ & Interleaver 2 & [$B$, $L$, $D_1$]  \\ 
        ~ & Component encoder 2 & [$B$, $L$, $D_2$]  \\ 
        ~ & Linear & [$B$, $L$, $D_2$]  \\ 
        \hline
        Channel & ~ & [$B$, $L$, 3$D_2$]  \\ \hline
         \multirow{8}{*}{Turbo decoder} & Interleaver 3 & [$B$, $L$, $D_2$] or [$B$, $L$, $D_4$]\\ 

        ~ & Forward decoder & [$B$, $L$, $D_3$]  \\ 
        ~ & Linear 1 & [$B$, $L$, $D_4$]  \\ 
        ~ & DeInterleaver 1 & [$B$, $L$, $D_4$]  \\ 
        ~ & Interleaver 4 &  [$B$, $L$, $D_2$] or [$B$, $L$, $D_4$] \\ 
        
        ~ & Backward decoder  & [$B$, $L$, $D_3$] or [$B$, $L$, $D_1$] \\ 
        ~ & Linear 2  & [$B$, $L$, $D_4$] or [$B$, $L$, $D_1$] \\ 
        ~ & DeInterleaver 2  & [$B$, $L$, $D_4$] or [$B$, $L$, $D_1$] \\ \hline
         \multirow{5}{*}{Semantic decoder} & Transformer decoder 1 & [$B$, $L$, $D_1$]  \\ 
        ~ & Transformer decoder 2 & [$B$, $L$, $D_1$]  \\ 
        ~ & Transformer decoder 3 & [$B$, $L$, $D_1$]  \\ 
        ~ & Linear & [$B$, $L$, $D_5$]  \\ 
        ~ & Softmax & [$B$, $L$, $D_5$]  \\ \hline
    \end{tabularx}
\end{table}

\subsection{Semantic Encoder}

The semantic encoder of Turbo-DSA is based on transformer, which deeply analyzes the input data and precisely extracts the core semantic features. Leveraging the attention mechanism of the transformer, it focuses on the key elements of the information, efficiently transforming the original messages into semantic vectors. During the training process, these semantic vectors are dynamically optimized according to the characteristics of the channel and the background knowledge base to adapt to various transmission conditions.

Specifically, the semantic encoder consists of $j$ transformer encoder layers \cite{DeepSC}. We transform the message $\bm{s}$ into a word representation vector ${\bm{s}}_{\rm{em}}$ through an embedding layer, serving as the input to the transformer encoder. The $j$-th transformer encoder layer consists of two sub-layers: multi-head self-attention and position-wise feed-forward networks.

The multi-head self-attention mechanism encompasses $h$ attention mechanisms, or heads, which concurrently calculate to capture the global relationships within the input sequence. It computes a representation for each position in the input sequence, considering information from other positions to capture various types of relationships and semantic information. The output of the multi-head self-attention mechanism is represented as
\begin{equation}
Z=f_{\rm{con}}({{\rm{Z}}_{\rm{1}}},\ldots ,{{\rm{Z}}_{\mathit{h}}}){{W}^{{\rm{o}}}} 
\end{equation}
where $f_{\rm{con}}(\bullet)$ denotes the concatenation function of multiple attention heads, ${{W}^{{\rm{o}}}}$ is an additional trainable weight matrix, and ${{\rm{Z}}_{i}}$ for the $i$-th attention head is computed as
\begin{equation}
{{Z}_{i}}={\rm{softmax(}}\frac{{{Q}_{{i}}}K_{{i}}^{\rm{T}}}{\sqrt{{{d}_{ki}}}}{\rm{)}}{{V}_{\mathit{i}}} 
\end{equation}
where $\rm{softmax}(\bullet)$ is the activation function, ${{d}_{ki}}$ is the dimensionality of $K_{i}$, and ${{Q}_{i}}$, $K_{i}$, ${{V}_{i}}$ are the queries, keys, and values for the $i$-th attention head, derived from the same input matrix ${\bm{s}_{\rm{em}}}$ through linear transformations
\begin{align}
 & Q_i={\bm{s}_{\rm{em}}}{{W}^{q_i}}  \\ 
 & K_i={\bm{s}_{\rm{em}}}{{W}^{k_i}}   \\ 
 & V_i={\bm{s}_{\rm{em}}}{{W}^{v_i}} 
\end{align}
where ${{W}^{q_i}}$, ${{W}^{k_i}}$, and ${{W}^{v_i}}$ are trainable parameter matrices.
After summing and normalizing the output $Z$, the input to the position-wise feed-forward networks, denoted as ${{Z}_{\rm{in}}}$, is obtained.

The feed-forward network consists of two linear layers and an activation function (e.g., ReLU). Applied independently at each position, it performs a non-linear transformation on each position in the sequence, enhancing the model's representational capacity. The output of the feed-forward network is given by
\begin{equation}
{{e}_{\rm{f}}}=\max (0,{{Z}_{\rm{in}}}{{W}_{\rm{1}}}+{{b}_{\rm{1}}}){{W}_{\rm{2}}}+{{b}_{\rm{2}}} 
\label{ffn}
\end{equation}
where ${{W}_{\rm{1}}}$, ${{W}_{\rm{2}}}$ are trainable weight matrices, and ${{b}_{\rm{1}}}$, ${{b}_{\rm{2}}}$ are trainable bias vectors. After summing and normalizing the output ${\rm{e}}_{\rm{f}}$, the output of the $j$-th transformer encoder layer, denoted as ${\rm{e}}_{j}$, is obtained, serving as the input for the next layer. This stacking of encoder layers facilitates the transmission and processing of information between layers.
After processing through $j$ transformer encoder layers, the output of the semantic encoder is denoted as $\mathbf{e}$.

In summary, the output signal of the semantic encoder of the proposed Turbo-DSA can be expressed as
\begin{equation}
\bm{e} = \emph{f}_{\rm{en}}^{\rm{\ trans}}(\bm{s}, \bm{\epsilon})
\label{semanticencoder}
\end{equation}
where $\emph{f}_{\rm{en}}^{\rm{\ trans}}$ represents the semantic encoding function, and $\bm{\epsilon}$ represents the parameters of the semantic encoder.

\subsection{Channel Encoder}

Subsequently, the channel encoder employs a DL-based Turbo structure to further encode the semantic vectors generated by the semantic encoder into a sequence of symbols. Redundancy is introduced through parallel concatenation, which introduces redundancy to the information, enhancing the fault tolerance of signal transmission. Turbo component encoding can separate and transmit semantic vectors, reducing interference between different vectors. This ensures accurate data transmission over maritime channels, maintaining high data integrity even under less than ideal channel conditions. In this way, Turbo-DSA optimizes the signal processing procedure, enhancing the stability and efficiency of communication in the marine environment.

As an efficient error correction coding technique, traditional Turbo encoders typically employ convolutional coding, utilizing recursive systematic convolutional encoders. In contrast, the DL-based Turbo encoder is implemented using CNN layers, capable of leveraging the advantages of big data processing to adaptively learn the characteristics of signals and optimize encoding strategies. Unlike the Ref. \cite{DEEPTURBO, jiang2019turbo}, the channel encoder of the proposed Turbo-DSA processes not only the data itself but also includes semantic information.

The structure of the channel encoder comprises three integral components: the main encoder, component encoder 1, and component encoder 2. This design allows Turbo encoding to fully leverage the independent redundant information from the two component encoders, providing more robust error correction capabilities to ensure the reliability of data transmission. The input signal for the DL-based Turbo encoder is a sequence of textual semantic features, denoted as signal $\bm{e}$. The DL-based Turbo encoder divides the input data $\bm{e}$ into multiple information blocks, categorized into two types. 

Firstly, the main encoder is responsible for generating the systematic sequence. The main encoder transforms the input signal $\bm{e}$ into the systematic sequence $\bm{x}^{\rm{sys}}$, which can be represented as
\begin{equation}
\bm{x}^{\rm{sys}} = \emph{f}_{\rm{en}}^{\rm{\ c}}(\bm{e}) 
\end{equation}
where $\emph{f}_{\rm{en}}^{\rm{\ c}}(\cdot)$ is the function of the main encoder. 

Secondly, both component encoder 1 and component encoder 2 are tasked with generating redundant sequence, respectively. These encoders improve their resilience to channel errors by rearranging the input sequence and altering the temporal relationships. 
After the input signal $\bm{e}$ is interleaved by interleaver 1, it is encoded by component encoder 1 to produce the sequence $\bm{x}^{\rm{par1}}$, which is represented as
\begin{equation}
\bm{x}^{\rm{par1}} = \emph{f}_{\rm{en}}^{\rm{\ int1}}(\bm{e}) 
\end{equation}
where $\emph{f}_{\rm{en}}^{\rm{\ int1}}(\cdot)$ is the function of component encoder 1. 
Component encoder 2 mirrors the structure of component encoder 1 but differs in the interleaver's sequence rearrangement, leading to a distinct set of redundant sequence. The output sequence from component encoder 2, $\bm{x}^{\rm{par2}}$ can be represented as
\begin{equation}
\bm{x}^{\rm{par2}} = \emph{f}_{\rm{en}}^{\rm{\ int2}}(\bm{e}) 
\end{equation}
where $\emph{f}_{\rm{en}}^{\rm{\ int2}}(\cdot)$ is the function of component encoder 2. 

Then, the outputs of the main encoder and component encoders are parallel concatenated to form the encoded sequence $\bm{x}$, which can be represented as
\begin{equation}
\bm{x} = [\bm{x}^{\rm{sys}}, \bm{x}^{\rm{par1}}, \bm{x}^{\rm{par2}}] 
\end{equation}

Therefore, the DL-based Turbo encoder enhances the system's fault tolerance and improves the transmission quality of communication signals in noisy channels. The signal from the channel encoder can be represented as
\begin{equation}
\bm{x} = \emph{f}_{\rm{en}}^{\rm{\ turbo}}(\bm{e}, \bm{\zeta})
\label{channelencoder}
\end{equation}
where $\emph{f}_{\rm{en}}^{\rm{\ turbo}}(\cdot)$ represents the channel encoding function, and $\bm{\zeta}$ represents the parameters of the channel encoder.

Subsequently, when transmitting signals over maritime channels, the signal is affected by fading channels and additive white Gaussian noise (AWGN). The signal transmitted over the channel can be represented as
\begin{equation}
\bm{r} = \bm{x} \bm{h} + \bm{n}
\label{channel}
\end{equation}
where $\bm{r}$ represents the signal received by the receiver, $\bm{n}$ represents the noise signal, and $\bm{h}$ represents the channel fading coefficients.

\subsection{Channel Decoder}

The channel decoder of Turbo-DSA employs a DL-based Turbo decoder, achieving a progressive enhancement in decoding accuracy through iterative decoding. Confronted with fluctuations in channel transmission, the Turbo iterative decoding dynamically adjusts the semantic vectors, effectively correcting errors that arise during the transmission process. At the same time, it optimizes and retains key elements within the semantic vectors, ensuring that the original intent and content of the information are accurately conveyed. This ensures that during the data transmission decoding phase, Turbo-DSA maintains the stability and integrity of semantic information under complex channel conditions.

Traditional Turbo decoders typically use two soft input soft output iterative decoders, employing multiple rounds of iterations to improve error correction performance. However, the DL-based Turbo decoder is implemented using CNN layers, which not only enhances decoding accuracy but also strengthens the ability to learn signal characteristics. Similar to the channel encoder, the channel decoder of Turbo-DSA differs from the Ref. \cite{DEEPTURBO, jiang2019turbo} in that it processes semantic information internally. During the decoding process, it not only focuses on the accurate recovery of data but, more importantly, achieves in-depth analysis and restoration at the semantic level.

The structure of the channel decoder consists of two identical iterative decoders, referred to as the forward decoder and the backward decoder. The Turbo decoder employs an iterative decoding, achieving multiple iterations by alternately running the forward and backward decoders.

Specifically, at the receiver end, the received signal $\bm{r}$ is divided into three types of sequences after serial-to-parallel conversion, namely the system information sequence $\bm{r}^{\rm{sys}}$, parity information sequence $\bm{r}^{\rm{par1}}$, and parity information sequence $\bm{r}^{\rm{par2}}$. They can be represented as 
\begin{align}
       & \bm{r}^{\rm{sys}}=[r^{\rm{sys}}_{\rm{1}}, r^{\rm{sys}}_{\rm{2}}, ... r^{\rm{sys}}_{\rm{n}}] \\ 
       & \bm{r}^{\rm{par1}}=[r^{\rm{par1}}_{\rm{1}}, r^{\rm{par1}}_{\rm{2}}, ... r^{\rm{par1}}_{\rm{n}}]   \\ 
       & \bm{r}^{\rm{par2}}=[r^{\rm{par2}}_{\rm{1}}, r^{\rm{par2}}_{\rm{2}}, ... r^{\rm{par2}}_{\rm{n}}]
\end{align}

Next, the system information sequence $\bm{r}^{\rm{sys}}$ is interleaved to generate the system information sequence $\bm{r}^{\rm{sys}}_{\rm{int1}}$. The a prior information sequence $\bm{r}^{\rm{prior}}_1$ is external information generated by the backward decoder and is subsequently deinterleaved. In the first iteration, $\bm{r}^{\rm{prior}}_1=0$. The a prior information sequence $\bm{r}^{\rm{prior}}_1$ is interleaved to generate the a prior information sequence $\bm{r}^{\rm{prior}}_{\rm{int1}}$. The a prior information sequence $\bm{r}^{\rm{prior}}_{\rm{int1}}$, the system information sequence $\bm{r}^{\rm{sys}}_{\rm{int1}}$, and the parity information sequence $\bm{r}^{\rm{par1}}$ are processed by the forward decoder to generate the sequence $\bm{r}^{\rm^{de1}}$, which can be represented as follows
\begin{equation}
\bm{r}^{\rm^{de1}} = \emph{f}_{\rm{de}}^{\rm^{\ 1}}(\bm{r}^{\rm{prior}}_{\rm{int1}}, \bm{r}^{\rm{sys}}_{\rm{int1}}, \bm{r}^{\rm{par1}}) 
\end{equation}
where $\emph{f}_{\rm{de}}^{\rm^{\ 1}}(\cdot)$ is the function of the forward decoder. 

Then, the sequence $\bm{r}^{\rm^{de1}}$ is processed by deinterleaver 1 and interleaver 4 to obtain a new a prior information sequence $\bm{r}^{\rm{prior}}_{\rm{int2}}$, which is independent of $\bm{r}^{\rm{sys}}_{\rm{int1}}$ and $\bm{r}^{\rm{par1}}$, and can be used as a prior information for the backward decoder. The system information sequence $\bm{r}^{\rm{sys}}$ is interleaved by interleaver 4 to generate the system information sequence $\bm{r}^{\rm{sys}}_{\rm{int2}}$. The a prior information sequence $\bm{r}^{\rm{prior}}_{\rm{int2}}$, the system information sequence $\bm{r}^{\rm{sys}}_{\rm{int2}}$, and the parity information sequence $\bm{r}^{\rm{par2}}$ are processed by the backward decoder to generate the sequence $\bm{r}^{\rm^{de2}}$, which can be represented as follows
\begin{equation}
\bm{r}^{\rm^{de2}} = \emph{f}_{\rm{de}}^{\rm^{\ 2}}(\bm{r}^{\rm{prior}}_{\rm{int2}}, \bm{r}^{\rm{sys}}_{\rm{int2}}, \bm{r}^{\rm{par2}}) 
\end{equation}
where $\emph{f}_{\rm{de}}^{\rm^{\ 2}}(\cdot)$ is the function of the backward decoder. The sequence $\bm{r}^{\rm^{de2}}$ is processed by deinterleaver 2, and the process is repeated. 

By continuously feeding back, interleaving, and deinterleaving, each iteration corrects the information obtained in the previous iteration and passes the corrected information to the next iteration. The Turbo decoder iteratively decodes the received signal, gradually correcting and recovering the textual semantic features. After multiple iterations, the Turbo decoder converges. If the set number of iterations is reached, the sequence $\bm{r}^{\rm^{de2}}$ is no longer iterated after passing through deinterleaver 2 and is directly output as the output signal $\bm{d}$. 

Thereby, the signal of the channel decoder can be represented as
\begin{equation}
    \bm{d} = \emph{f}_{\rm{de}}^{\rm{\ turbo}}(\bm{r}, \bm{\eta})
    \label{channeldecoder}
\end{equation}
where $\emph{f}_{\rm{de}}^{\rm{\ turbo}}(\cdot)$ represents the channel decoding function, and $\bm{\eta}$ represents the parameters of the channel decoder.

\subsection{Semantic Decoder}

Similar to the semantic encoder, the semantic decoder is also constructed based on transformer, capable of efficiently decoding the received semantic vectors. Utilizing the attention mechanism, the transformer can deeply understand the context, accurately restore semantic information, and convert it back to the original message. This process completes the accurate transmission of information throughout the communication process, maintaining efficient semantic communication even in complex communication environments.

Specifically, the semantic decoder of the proposed Turbo-DSA is also composed of stacked transformer decoder layers \cite{DeepSC}. Each decoder layer consists of self-attention mechanism, encoder-decoder attention mechanism and position-wise feedforward neural network. The self-attention mechanism is used to compute representations based on the decoder's own output sequence, capturing and integrating internal dependencies within the sequence, providing rich contextual information for decoding. The encoder-decoder attention mechanism is used to align the encoder's output sequence with the decoder's current position, ensuring the decoder can accurately track the semantic features of the encoder and achieve effective information integration. The structure of the position-wise feedforward neural network is the same as that in the encoder, further refining the semantic representation of each position and enhancing the decoder's contextual understanding of each word in the sequence.

The output is a conditional probability distribution for generating the next target word based on the preceding context and the current partially generated target sequence. The signal after the semantic decoder can be represented as
\begin{equation}
\bm{p} = \emph{f}_{\rm{de}}^{\rm{\ trans}}(\bm{d}, \bm{\kappa})
\label{semanticdecoder}
\end{equation}
where $\emph{f}_{\rm{de}}^{\rm{\ trans}}(\cdot)$ represents the decoder function and $\bm{\kappa}$ represents the parameters of the semantic decoder.

\subsection{Loss Function}

In the design of Turbo-DSA, our objective is to minimize the semantic disparity between predictions and reference sentences through E2E training and maximize the rate of semantic comprehension. We employ the cross-entropy loss function to calculate the difference between predicted values and ground truth for each word, summing these differences for all words to obtain the total loss. Specifically, the loss function can be represented by the formula
\begin{equation}
L = - \frac{1}{N}\sum\limits_{i = 1}^N {\frac{1}{{L_i}}} \sum\limits_{j = 1}^{L_i} {y_{i,j}} \log ({\hat y_{i,j}} + \delta )
\label{loss}
\end{equation}
where $i$ represents the index of the estimated sentence $\bm{y}$, $N$ represents the number of samples, ${L_i}$ represents the length of the $i$-th sample, ${y_{i,j}}$ denotes the true value of the $j$-th label in the $i$-th sample, ${\hat y_{i,j}}$ represents the predicted value of the $j$-th label in the $i$-th sample and $\delta$ is a very small constant used to prevent division by zero when calculating $\log(0)$.

\section{Algorithm Design}
\label{Algorithm}

\subsection{Training Algorithm for Turbo-DSA}
Algorithm \ref{Training} outlines the training process for Turbo-DSA. Turbo-DSA employs an E2E training approach, ensuring consistency and coordination in the system's learning and optimization process. By optimizing the objective function, Turbo-DSA focuses on precise semantic extraction and recovery, achieving rapid convergence through an iterative optimization process.
The input is the training data set $\bm{s}$. We initialize the parameter $i$ and commence iterative training using gradient descent, continuing until the loss function converges. Each iteration involves the following steps.
The transmitter encodes the signal $\bm{s}$ into $\bm{e}$ via the semantic encoder and subsequently into $\bm{x}$ through the channel encoder.
The signal $\bm{x}$ is transmitted over the channel, where it is subject to the maritime channel's specific parameters $\bm{h}$ and noise $\bm{n}$, resulting in the received signal $\bm{r}$.
In the receiver, the received signal $\bm{r}$ is decoded back into $\bm{d}$ by the channel decoder and then into $\bm{p}$ by the semantic decoder, yielding the output signal $\bm{p}$.
The loss function, given by equation (\ref{loss}), is calculated based on the discrepancy between the input signal $\bm{s}$ and the output signal $\bm{p}$.
Parameters are fine-tuned using gradient descent to minimize the loss.
Upon convergence, the trained Turbo-DSA network parameters $\bm{\epsilon}$, $\bm{\zeta}$, $\bm{\eta}$, $\bm{\kappa}$ are stored, marking the completion of the training phase.

\begin{algorithm}[t]
\caption{\textbf{Training Algorithm for Turbo-DSA}}
\label{Training}
\begin{algorithmic}[1] 
\State \textbf{Input:} Training data set $\bm{s}$.
\State \textbf{Output:} Turbo-DSA network parameters $\bm{\epsilon}$, $\bm{\zeta}$, $\bm{\eta}$, $\bm{\kappa}$.
\Statex 
\State Initialize $i = 0$.
\While{Loss function not converged}
    \State \textbf{Encoder:} 
    \State \qquad $\bm{e}$ $\gets$ $\emph{f}_{\rm{en}}^{\rm{\ trans}}(\bm{s}, \bm{\epsilon})$
    \State \qquad $\bm{x}$ $\gets$ $\emph{f}_{\rm{en}}^{\rm{\ turbo}}(\bm{e}, \bm{\zeta})$
    \State \qquad Transmit signal $\bm{x}$
    
    \State \textbf{Maritime Channel:}
    \State \qquad $\bm{y}$ $\gets$ $\bm{r} \bm{h} + \bm{n}$
    
    \State \textbf{Decoder:}
    \State \qquad Receive signal $\bm{r}$.
    \State \qquad $\bm{d}$ $\gets$ $\emph{f}_{\rm{de}}^{\rm{\ turbo}}(\bm{r}, \bm{\eta})$
    \State \qquad $\bm{p}$ $\gets$ $\emph{f}_{\rm{de}}^{\rm{\ trans}}(\bm{d}, \bm{\kappa})$
    
    \State Calculate the loss function $L$ using (\ref{loss}) based on the input signal $\bm{s}$ and the output signal $\bm{p}$.
    \State Optimize the training using gradient descent.
    \State Update parameters $\bm{\epsilon}$, $\bm{\zeta}$, $\bm{\eta}$, $\bm{\kappa}$.
    \State $i$ $\gets$ $i+1$
\EndWhile
\State Save the trained parameters.
\end{algorithmic}
\end{algorithm}

\subsection{Testing Algorithm for Turbo-DSA}

Algorithm \ref{Testing} details the testing process for Turbo-DSA. The testing algorithm conducts a comprehensive assessment of Turbo-DSA's performance under various SNR conditions.
The input comprises the test data set $\bm{s}'$, the trained Turbo-DSA network parameters $\bm{\epsilon}$, $\bm{\zeta}$, $\bm{\eta}$, $\bm{\kappa}$, and a spectrum of SNR values. The goal is to produce output signals $\bm{p}'$ under various SNR conditions. The testing procedure for each SNR value includes the following steps.
Encoding the test signal $\bm{s}'$ into $\bm{e}'$ through the semantic encoder and then into $\bm{x}'$ via the channel encoder.
Transmitting the signal $\bm{x}'$ over the channel, where it experiences the maritime channel effects consistent with the training phase, leading to the received signal $\bm{r}'$.
Decoding the received signal $\bm{r}'$ into $\bm{d}'$ by the channel decoder and subsequently into $\bm{p}'$ by the semantic decoder, generating output signals $\bm{p}'$ for the given SNR value.
This testing methodology ensures a thorough evaluation of the system's performance across different channel conditions, providing insights into the system's effectiveness.

\begin{algorithm}[t]
\caption{\textbf{Testing Algorithm for Turbo-DSA}}
\label{Testing}
\begin{algorithmic}[1] 
\State \textbf{Input:} Test data set $\bm{s}'$, Turbo-DSA network parameters $\bm{\epsilon}$, $\bm{\zeta}$, $\bm{\eta}$, $\bm{\kappa}$, Range of SNR values.
\State \textbf{Output:} Output signals $\bm{p}'$ at different SNR values.
\Statex 
\For{Each SNR value}
 \State \textbf{Encoder:} 
    \State \qquad $\bm{e}'$ $\gets$ $\emph{f}_{\rm{en}}^{\rm{\ trans}}(\bm{s}', \bm{\epsilon})$
    \State \qquad $\bm{x}'$ $\gets$ $\emph{f}_{\rm{en}}^{\rm{\ turbo}}(\bm{e}', \bm{\zeta})$
    \State \qquad Transmit signal $\bm{x}$
    
    \State \textbf{Maritime Channel:}
    \State \qquad $\bm{r}'$ $\gets$ $\bm{x}' \bm{h} + \bm{n}$
    
    \State \textbf{Decoder:}
    \State \qquad Receive signal $\bm{r}'$.
    \State \qquad $\bm{d}'$ $\gets$ $\emph{f}_{\rm{de}}^{\rm{\ turbo}}(\bm{r}', \bm{\eta})$
    \State \qquad $\bm{p}'$ $\gets$ $\emph{f}_{\rm{de}}^{\rm{\ trans}}(\bm{d}', \bm{\kappa})$
\EndFor
\end{algorithmic}
\end{algorithm}

\section{Simulation Experiments}
\label{Simulation}

This section details the configuration of simulation experiments and the performance evaluation. The simulation settings encompass essential parameter settings, the datasets employed, the simulated channel environments, the benchmark systems utilized for comparative analysis, and the array of metrics deployed to assess performance. The performance analysis is dedicated to presenting findings from simulation experiments that were conducted across a variety of channel conditions, employing different datasets, exploring diverse network architectures, and examining various combinations of training parameters.

\subsection{Simulation Settings}

\subsubsection{Parameter Settings}

The training parameters for the Turbo-DSA are detailed in Tab. \ref{setting}. We utilize the Adam optimizer, initiating the learning rate at 0.001. For assessing network performance, we incorporate metrics such as Bilingual Evaluation Understudy (BLEU) and Sentence Similarity (SS). The SNR, which is commonly ascertained at the receiver, is designated at a training value of 4 dB. During the training phase, the batch size is configured to 128, and the Rician factor $K$ is established at 3.

\begin{table}[h]
    \centering
    \caption{The training parameters of the proposed Turbo-DSA.}\label{setting}
    \begin{tabularx}{0.8\textwidth}{>{\centering\arraybackslash}p{0.4\textwidth} >{\centering\arraybackslash}p{0.4\textwidth}}
    \hline
    \textbf{Parameter name} & \textbf{Settings} \\
    \hline
        Optimizer & Adam \\
        Network performance indicator & BLEU, SS\\
        Initial learning rate & 0.0001 \\
        Rician factor $K$ & 3 \\
        Training SNR & 2 dB \\
        Batch size $B$ & 128 \\
        Sentence length $L$ & 30 \\
        Dimension $D_1$ & 128 \\
        Dimension $D_2$ & 16 \\
        Dimension $D_3$ & 100 \\
        Dimension $D_4$ & 5 \\
        Dimension $D_5$ & 35632 \\
        \hline
    \end{tabularx}
\end{table}

\subsubsection{Datasets}

In maritime communications, the transmission of data packets is typically bifurcated into two components: the header and the payload. As this study is dedicated to assessing the efficacy of Turbo-DSA in semantic extraction, we concentrate on the data segments within the payload, especially those encompassing maritime textual content. To maximize the potential of Turbo-DSA for semantic extraction in maritime communication networks, the datasets employed in this research encompass both widely recognized datasets for textual semantic communication systems and those enriched with maritime background knowledge. The datasets are as follows.

\begin{enumerate}
    \item General dataset: The European Parliament Proceedings Parallel Corpus, renowned for its extensive scale and multi-lingual content, encompassing a diverse array of global development topics such as politics, economy, society, culture, and human rights \footnote{https://www.statmt.org/europarl/}.
    \item Maritime dataset 1: The book "Next Generation Marine Wireless Communication Networks," published by Springer \cite{BinLinBook}, offering a comprehensive overview of next-generation maritime wireless communication networks, marine economy and safety, marine industry, marine tourism, marine monitoring systems, and the GMDSS, among other pertinent fields. 
    \item Maritime dataset 2: This dataset comprises a collection of documents sourced from the International Maritime Organization (IMO) official website, primarily focusing on the IMO terminology database, guidelines for ship recycling plan development, authorization of ship recycling facilities, safe and environmentally sound ship recycling practices, and standard maritime communication protocols, among other relevant areas \footnote{https://www.imo.org/}. 
\end{enumerate}

\subsubsection{Channels}

To evaluate the performance of Turbo-DSA across a range of marine communication environments, from ideal to highly variable, we select channel models: AWGN, Rician fading, and Rayleigh fading channels.

\begin{enumerate}
    \item The AWGN channel, defined by its simple Gaussian noise profile, provides an idealized framework for establishing the baseline performance of Turbo-DSA. It serves as a pristine benchmark for assessing the capabilities of marine IoT communication links, particularly in the absence of multipath interference and shadow fading.
    \item The Rician fading channel models the integration of line-of-sight and non-line-of-sight propagation modes, reflecting the practical challenges in communications between offshore platforms and near-shore devices \cite{riciank,lee2017measurement,wang2018wireless,yu2020maritime}. It offers critical insights for the design of resilient and robust maritime communication systems by accurately depicting the coexistence of direct line paths and multipath reflections. Mathematically, this channel can be represented as
\begin{equation}
    h = \sqrt{\frac{K}{K+1}}\cdot e^{j\phi} + \sqrt{\frac{1}{K+1}}\cdot n
\end{equation}
    where, the Rician factor $K$ governs the ratio between the direct and reflected signals, while $n$, a random variable, encapsulates the uncertainty inherent in multipath signals.
    \item The Rayleigh fading channel highlights the dynamic nature of multipath propagation, particularly in environments lacking a direct line of sight. By simulating the fluctuations in signal strength due to multiple reflections and scatterings, it reveals the unique attenuation patterns characteristic of oceanic environments.
\end{enumerate}

\subsubsection{Benchmarks}

To compare the performance of Turbo-DSA, we choose the following benchmarks.

\begin{enumerate}
    \item CNN-AE: A communication system incorporates CNN layers within its encoder and decoder \cite{CNN-AE}.
    \item DeepSC: A semantic communication framework, DeepSC is characterized by its use of dense layers for channel encoder and decoder \cite{DeepSC}.
    \item DSA: A semantic communication system, DSA is distinguished by its use of CNN layers for channel encoding and decoding \cite{DSA}. 
\end{enumerate}


\subsubsection{Performance Metrics}

To evaluate the performance of Turbo-DSA in semantic transmission, we employ the key performance metrics of BLEU and SS to analyze its effectiveness in handling semantic information.


\begin{enumerate}
    \item BLEU: Designed to measure the consistency between the input and received sentences \cite{DeepSC}.
    It operates on the precision and recall of n-grams, with a weighted mechanism that evaluates the degree of phrase overlap. For example, the sentence "IMO is the only United Nations specialized agency with its headquarters in the United Kingdom" can be analyzed through 1-grams (e.g., "IMO", "is", "the"), 2-grams (e.g., "IMO is", "is the", "the only"), 3-grams (e.g., "IMO is the", "is the only", "the only United"), and 4-grams (e.g., "IMO is the only", "is the only United", "the only United Nations"). The BLEU score can be expressed as
\begin{equation}
    \text{BLEU} = BP \exp\left(\sum_{n=1}^{N} w_n \log p_n\right)
\end{equation}
    where $BP $ is the brevity penalty, $w_n $ denotes the weight assigned to each n-gram (commonly equal), and $p_n $ reflects the precision of n-grams matched between the candidate and reference sentences.

    \item SS: As a metric for assessing the similarity between two sentences \cite{zhishitupu}, the process begins with sentence tokenization and the mapping of words to integers.
    The BERT model then encodes the input and output sentences into vectors $ h_{\rm{input}} $ and $ h_{\rm{output}} $, respectively. The SS can be expressed as
\begin{equation}
    \text{SS} = \frac{h_{\rm{input}} \cdot h_{\rm{output}}}{\left\| h_{\rm{input}} \right\| \cdot \left\| h_{\rm{output}} \right\|}
\end{equation}

\end{enumerate}

\subsection{Performance Analysis}

\subsubsection{Maritime Dataset Analysis}

\begin{figure} 

    \centering
        \subfloat[General dataset]{\label{Simu_DataEuropeSmall_Rayleigh_BLEU}\includegraphics[width=0.9\linewidth]{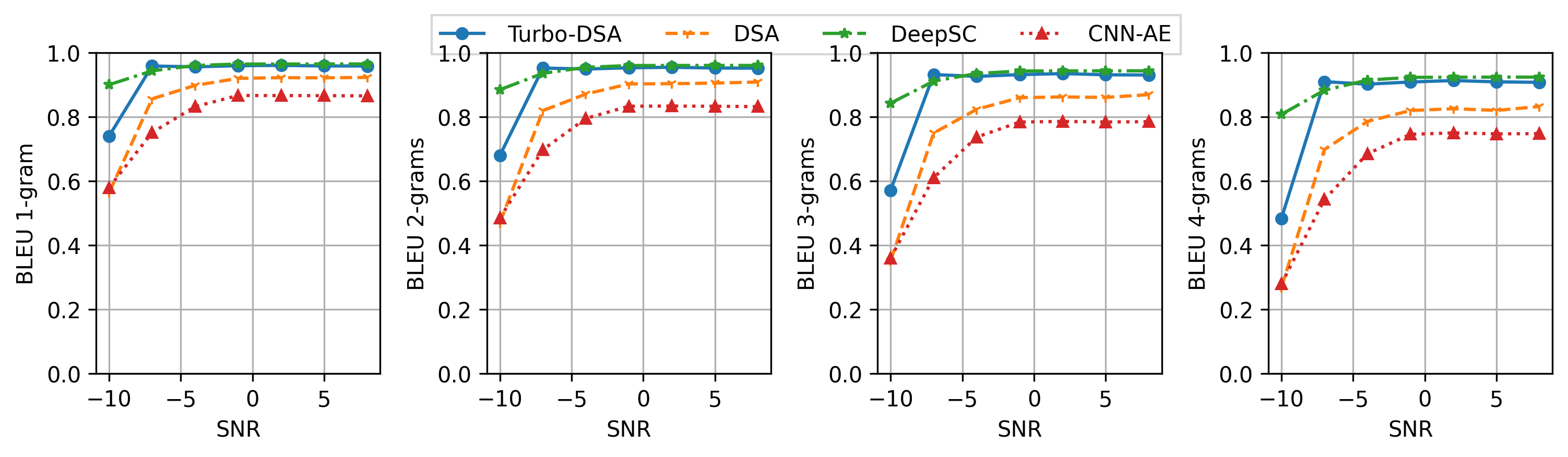}}
        
        \subfloat[Maritime dataset 1]{\label{Simu_DataBook_Rayleigh_BLEU}\includegraphics[width=0.9\linewidth]{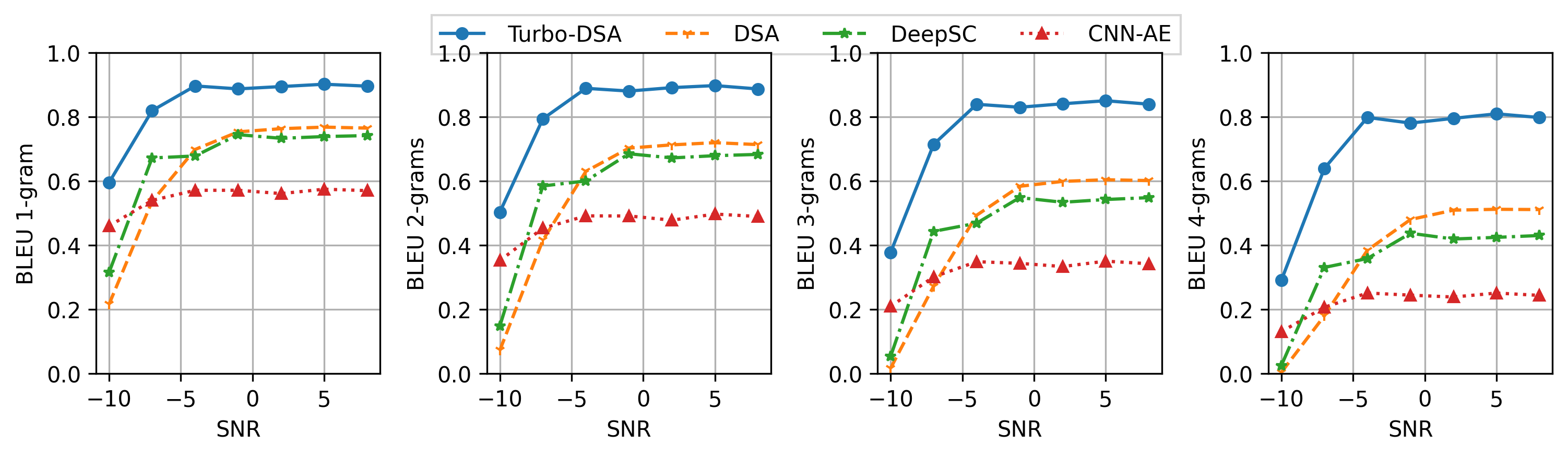}}
        
        \subfloat[Maritime dataset 2]{\label{Simu_DataIMO_Rayleigh_BLEU}\includegraphics[width=0.9\linewidth]{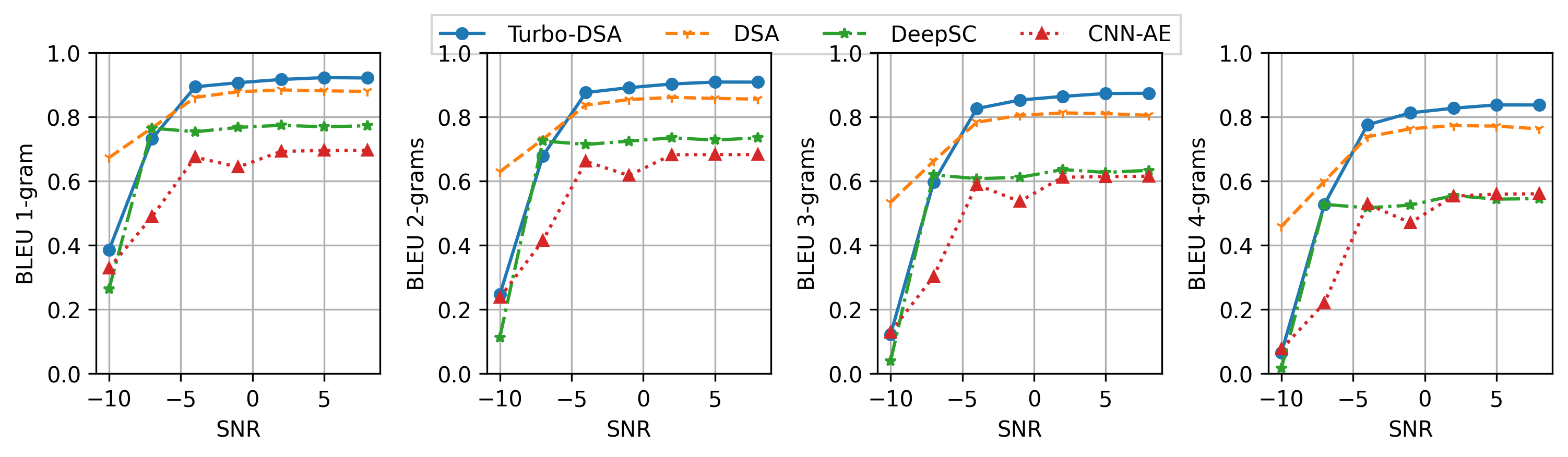}}
        
        \caption{The BLEU scores versus SNR under different datasets.}
        \label{DatasetBLEUK}

\end{figure}
\begin{figure} 
        \centering
        \subfloat[General dataset]{\label{Simu_DataEuropeSmall_Rayleigh_RayleighSS}\includegraphics[width=0.3\linewidth]{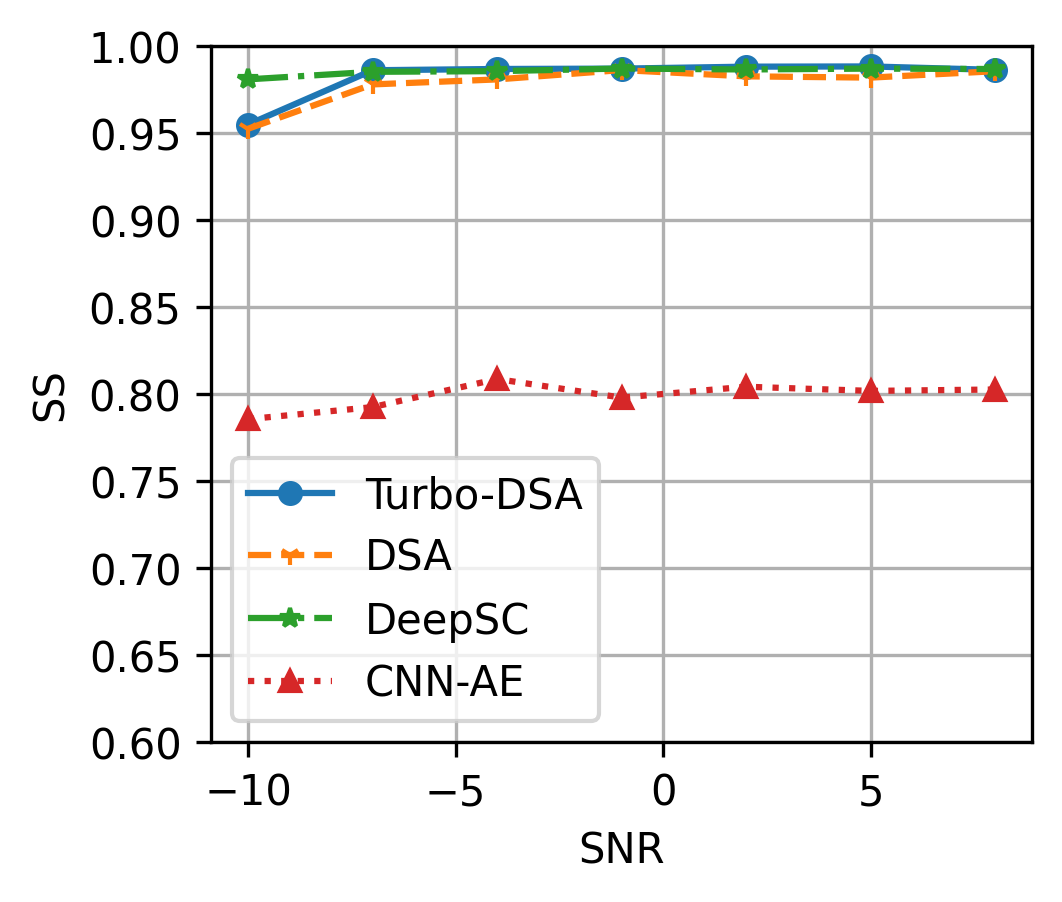}}
        \subfloat[Maritime dataset 1]{\label{Simu_DataBook_Rayleigh_RayleighSS}\includegraphics[width=0.3\linewidth]{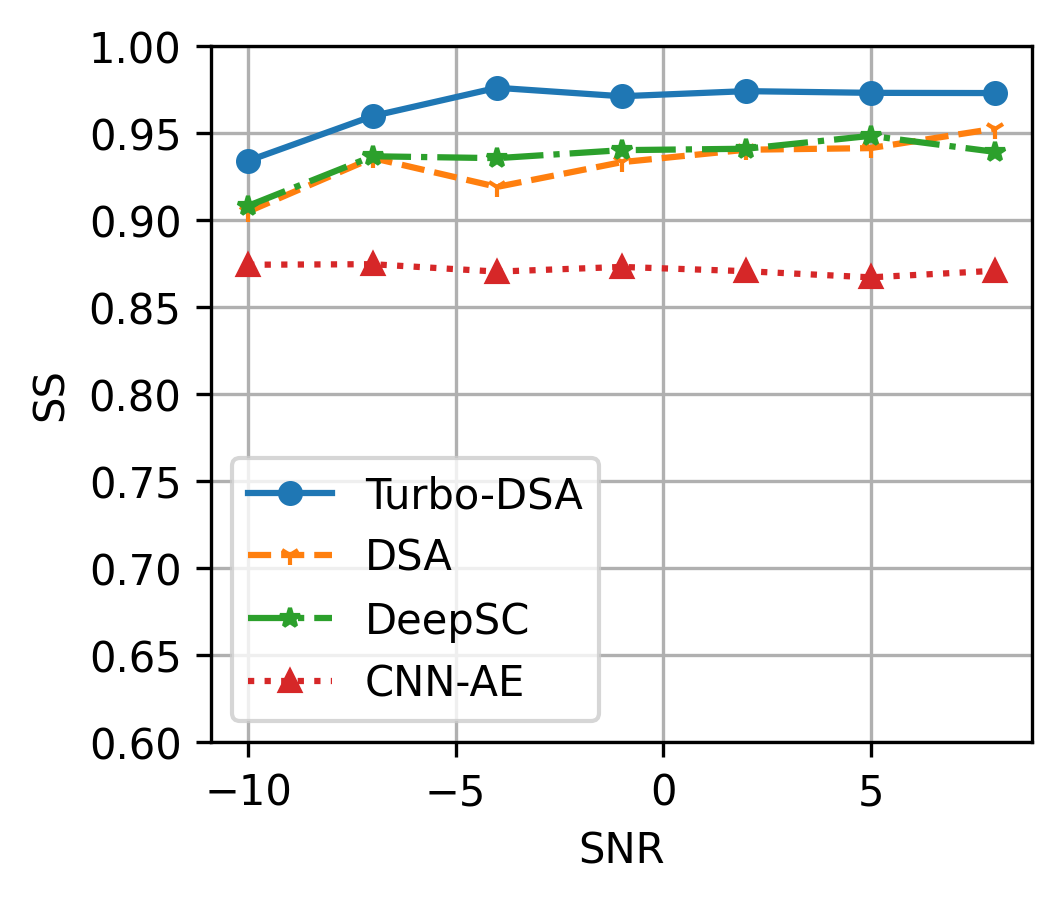}}
        \subfloat[Maritime dataset 2]{\label{Simu_DataIMO_Rayleigh_RayleighSS}\includegraphics[width=0.3\linewidth]{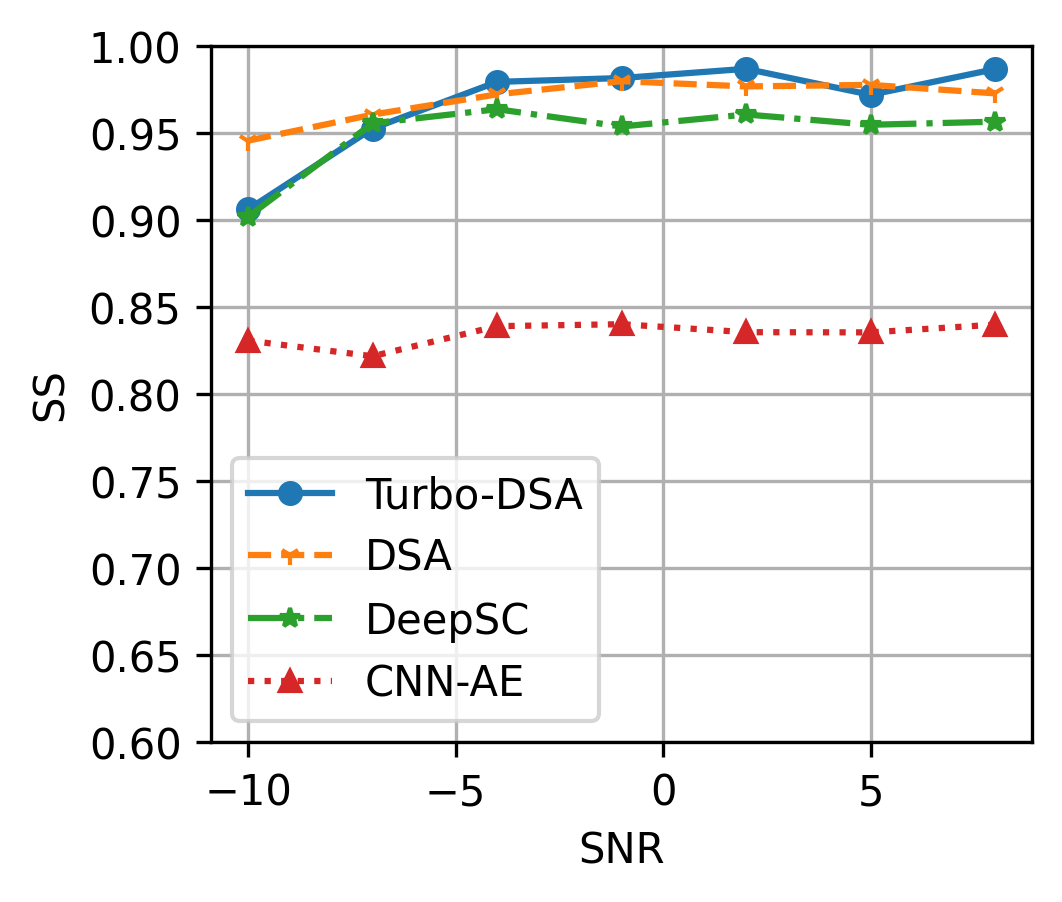}}
        \caption{The SS versus SNR under different datasets.}
        \label{DatasetSS}
\end{figure}

To comprehensively evaluate the performance of Turbo-DSA across different datasets, we conduct targeted simulation experiments. As revealed in Fig. \ref{DatasetBLEUK} and Fig. \ref{DatasetSS}, Turbo-DSA demonstrated exceptional performance across all datasets, with its adaptability particularly evident in the maritime IoT environment.

On general datasets, as shown in Fig. \ref{Simu_DataEuropeSmall_Rayleigh_BLEU} and Fig. \ref{Simu_DataEuropeSmall_Rayleigh_RayleighSS}, Turbo-DSA, like other comparative systems, exhibits excellent BLEU and SS performance. This is because general datasets have a wide coverage and a rich knowledge base, enabling the model to effectively extract sentence semantics by learning a large amount of text.

%
%
%
%
%

However, the performance advantage of Turbo-DSA is more significant in maritime dataset 1 and maritime dataset 2. Maritime dataset 1 covers multiple aspects of the maritime industry, as shown in Fig. \ref{Simu_DataBook_Rayleigh_BLEU} and Fig. \ref{Simu_DataBook_Rayleigh_RayleighSS}. When trained on maritime dataset 1, Turbo-DSA significantly outperforms other comparative systems in BLEU and SS metrics. The specific domain nature of maritime dataset 1 makes the sentences more professional, and Turbo-DSA's professionalism allows it to effectively extract maritime text semantics.
Maritime dataset 2 is even more focused on maritime terminology databases, ship recycling, and other specialized knowledge. As shown in Fig. \ref{Simu_DataIMO_Rayleigh_BLEU} and Fig. \ref{Simu_DataIMO_Rayleigh_RayleighSS}, even on such a specialized dataset, Turbo-DSA still leads other systems in BLEU and SS performance, further demonstrating its superiority in maritime text semantics transmission.

The experimental results clearly indicate that when the dataset is large in scale and has a wide range of knowledge coverage, all models can exhibit good performance, thanks to the powerful learning capabilities of structures such as the transformer. However, when the dataset becomes highly specialized and focused, especially when it possesses distinct industry characteristics like maritime data, Turbo-DSA, due to its unique component encoding and iterative decoding mechanism, is able to maintain stable performance in complex and variable channel environments, effectively adapting to the knowledge system of professional fields. This not only reinforces the superiority of Turbo-DSA in maritime text semantics transmission but also reflects its strong adaptability to variable channel environments, making it a trustworthy choice in the field of maritime communication.

\subsubsection{Wireless Communication Channel Analysis}

\begin{figure*}

    \centering
        \subfloat[AWGN]{\label{Simu_Comp_AWGN_BLEU}\includegraphics[width=0.9\linewidth]{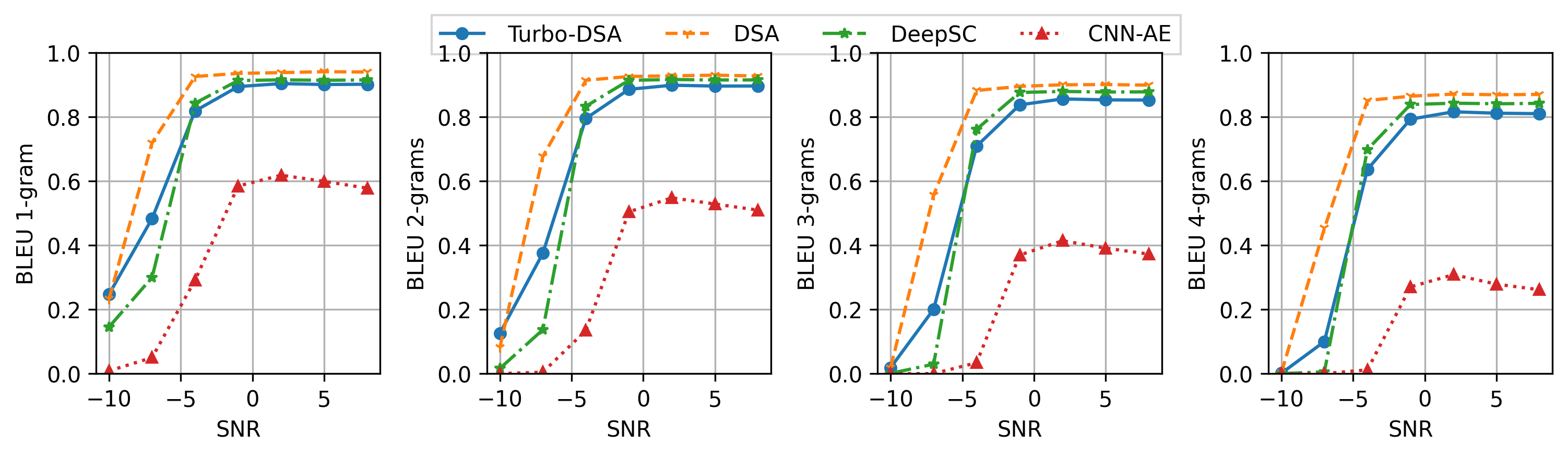}}
        
        \subfloat[Rician]{\label{Simu_Comp_newnewRician_BLEU}\includegraphics[width=0.9\linewidth]{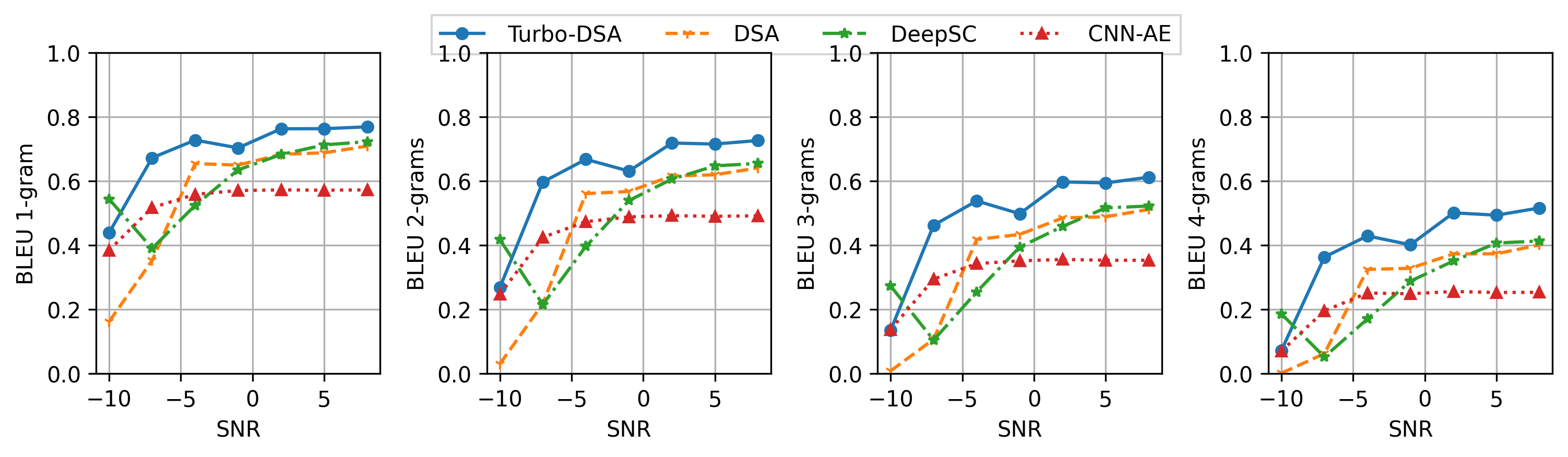}}
        
        \subfloat[Rayleigh]{\label{Simu_Comp_Rayleigh_BLEU}\includegraphics[width=0.9\linewidth]{figs/Simu_Comp_Rayleigh_BLEU}}

        \caption{The BLEU scores versus SNR under different channels, using the maritime dataset 1.}
        \label{channelBLEUK}
\end{figure*}
    
\begin{figure*}

    \centering
        \subfloat[AWGN]{\label{Simu_Comp_AWGN_AWGNSS}\includegraphics[width=0.3\linewidth]{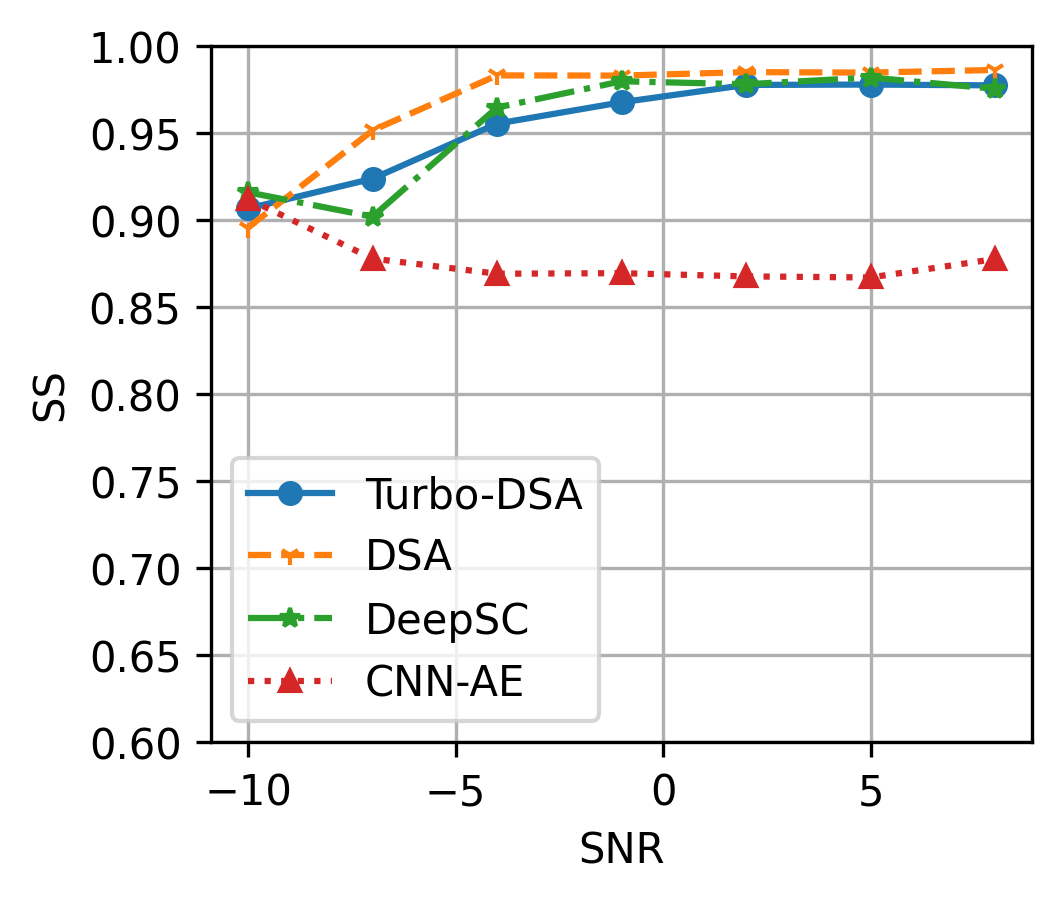}}
        \subfloat[Rician]{\label{Simu_Comp_newnewRician_newnewRicianSS}\includegraphics[width=0.3\linewidth]{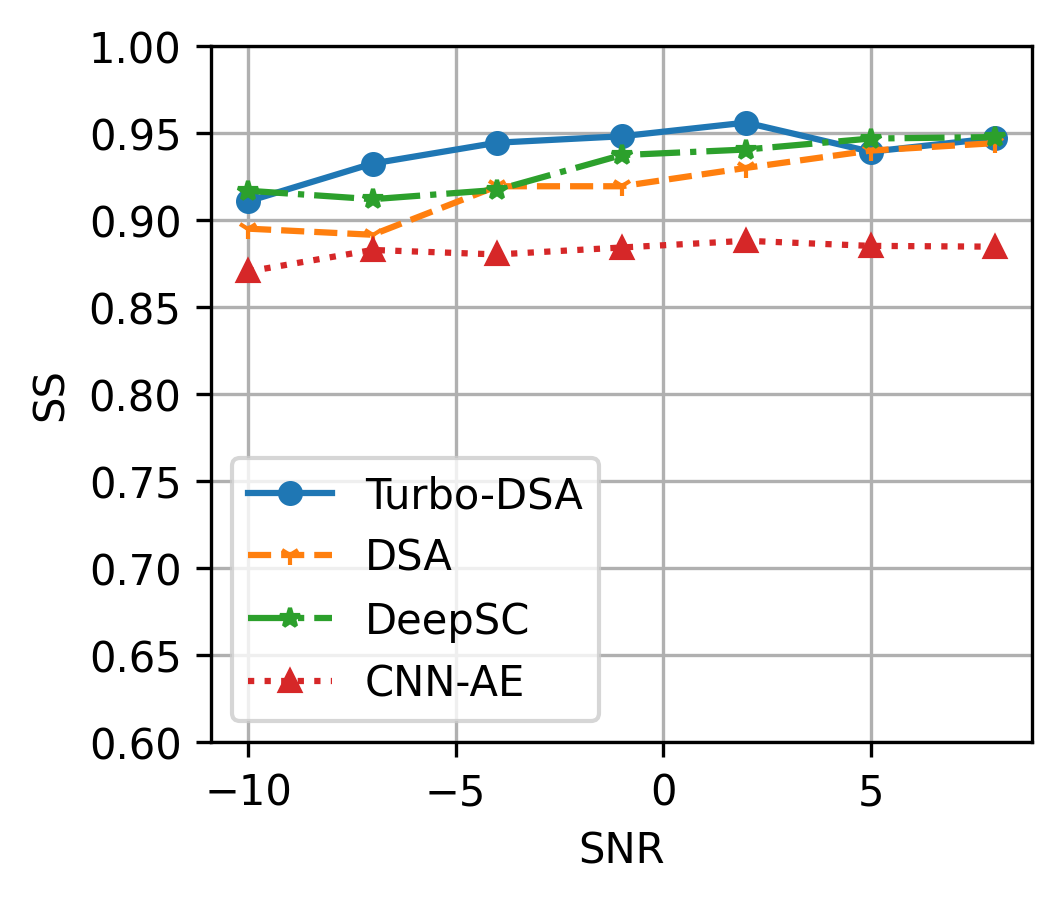}}
        \subfloat[Rayleigh]{\label{Simu_Comp_Rayleigh_RayleighSS}\includegraphics[width=0.3\linewidth]{figs/Simu_Comp_Rayleigh_RayleighSS}}

        \caption{The SS versus SNR under different channels, using the maritime dataset 1.}
        \label{channelSS}

\end{figure*}

To thoroughly compare the performance of Turbo-DSA in various wireless communication environments, we conduct a series of simulation experiments.
Fig. \ref{channelBLEUK} and Fig. \ref{channelSS} display the maritime text semantic transmission performance of Turbo-DSA under different wireless communication channel conditions. 
Due to the dynamic and variable nature of the marine environment, we tested the system's performance in various common wireless communication channel models, including Gaussian white noise channels, Ricean fading channels, and Rayleigh fading channels, to analyze the model's performance capabilities under different channel conditions.

In the most ideal channel conditions, namely the Gaussian white noise channel, as shown in Fig. \ref{Simu_Comp_AWGN_BLEU} and Fig. \ref{Simu_Comp_AWGN_AWGNSS}, Turbo-DSA, like other comparative systems, exhibits excellent BLEU and SS performance. This is because, under such ideal channel conditions, the model can accurately capture and convey semantic information through learning a large amount of text.

However, when the channel conditions deteriorate into Ricean fading channels, the performance advantage of Turbo-DSA begins to stand out. Ricean fading channels contain a line-of-sight link, which closely aligns with the real-world scenarios of maritime communication. As shown in Fig. \ref{Simu_Comp_newnewRician_BLEU} and Fig. \ref{Simu_Comp_newnewRician_newnewRicianSS}, in Ricean fading channels, Turbo-DSA significantly outperforms other comparative systems in BLEU and SS metrics. This indicates that Turbo-DSA can adapt to this channel environment, effectively extracting the semantic information of maritime text. Its coding technology ensures efficient extraction and transmission of maritime text semantics, even in the presence of a line-of-sight link.

Maritime wireless communication environments are highly changeable, and the line-of-sight link may suddenly be interrupted due to the movement of vessels or the influence of waves. Therefore, we further considered the scenario without a line-of-sight link, simulating the situation using Rayleigh fading channels. As shown in Fig. \ref{Simu_Comp_Rayleigh_BLEU} and Fig. \ref{Simu_Comp_Rayleigh_RayleighSS}, even in wireless channel environments lacking a line-of-sight link, Turbo-DSA still leads other systems in BLEU and SS performance, further demonstrating its superiority in maritime text semantics transmission.

The experimental results clearly reveal that in ideal wireless communication environments, all models can perform well. However, once the channel conditions become complex and variable, Turbo-DSA, with its innovative component encoding and iterative decoding strategy, demonstrates excellent adaptability and stability. These findings not only confirm the great potential of Turbo-DSA in the field of maritime communication but also emphasize its reliability in facing unpredictable channel challenges in real-world applications. The success of Turbo-DSA proves its advanced and practical design in semantic communication systems, laying a solid foundation for the development of future maritime communication technologies.

\subsubsection{Rician Fading Channel Analysis}

\begin{figure*}
        \centering
        \subfloat[$K=1$]{\label{Simu_DataIMO_newnewRician_BLEU_K1}\includegraphics[width=0.9\linewidth]{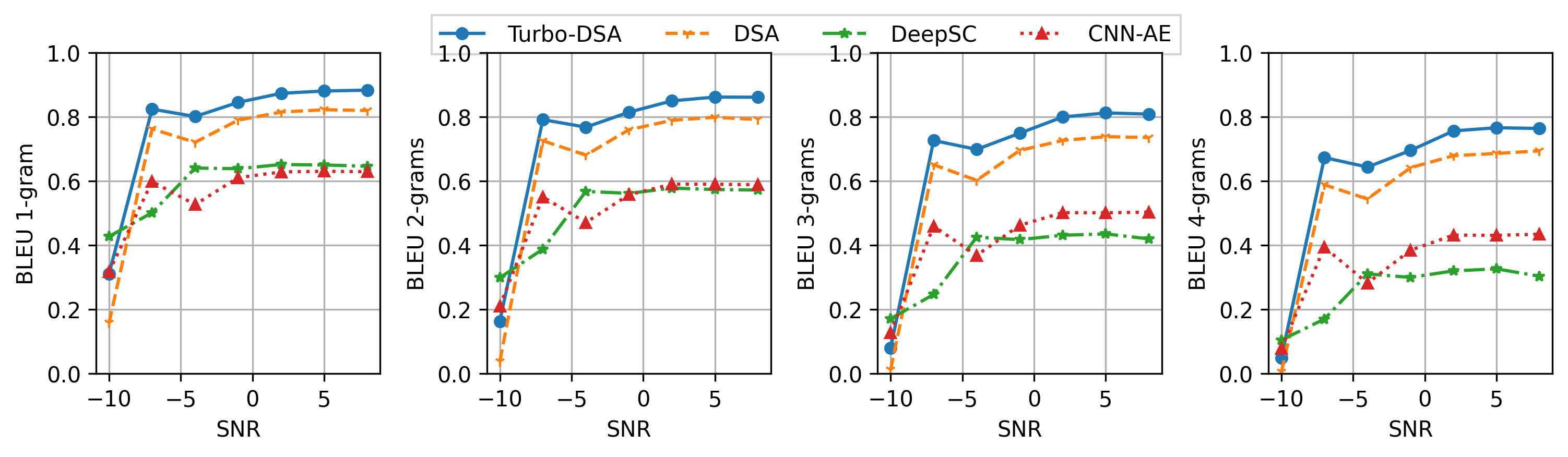}}
        
        \subfloat[$K=4$]{\label{Simu_DataIMO_newnewRician_BLEU_K4}\includegraphics[width=0.9\linewidth]{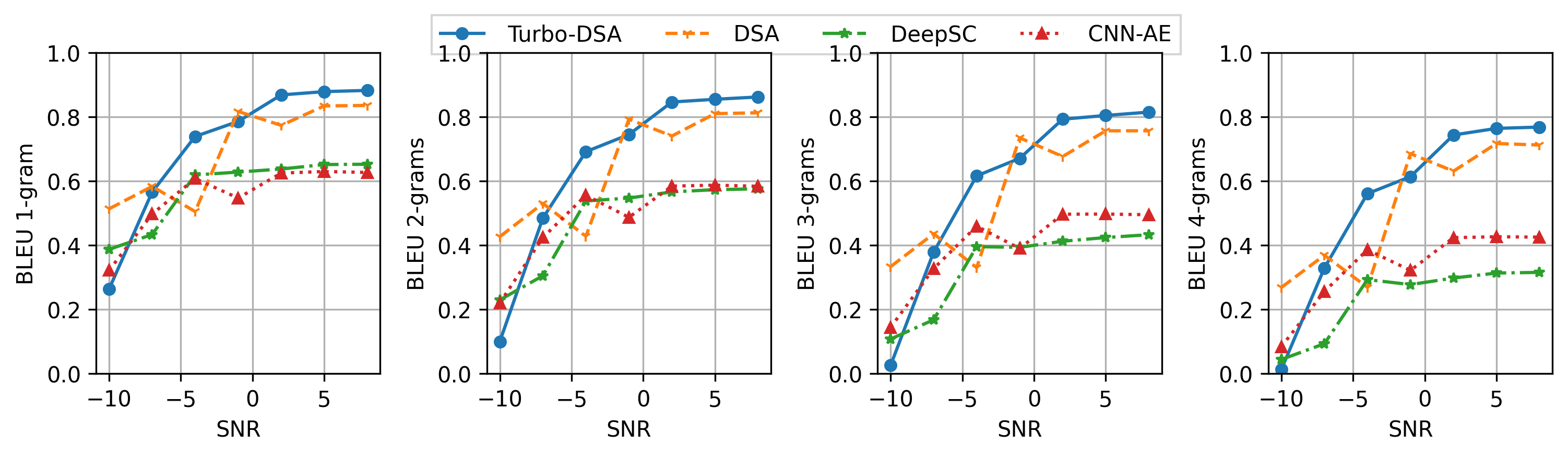}}
        
        \subfloat[$K=7$]{\label{Simu_DataIMO_newnewRician_BLEU_K7}\includegraphics[width=0.9\linewidth]{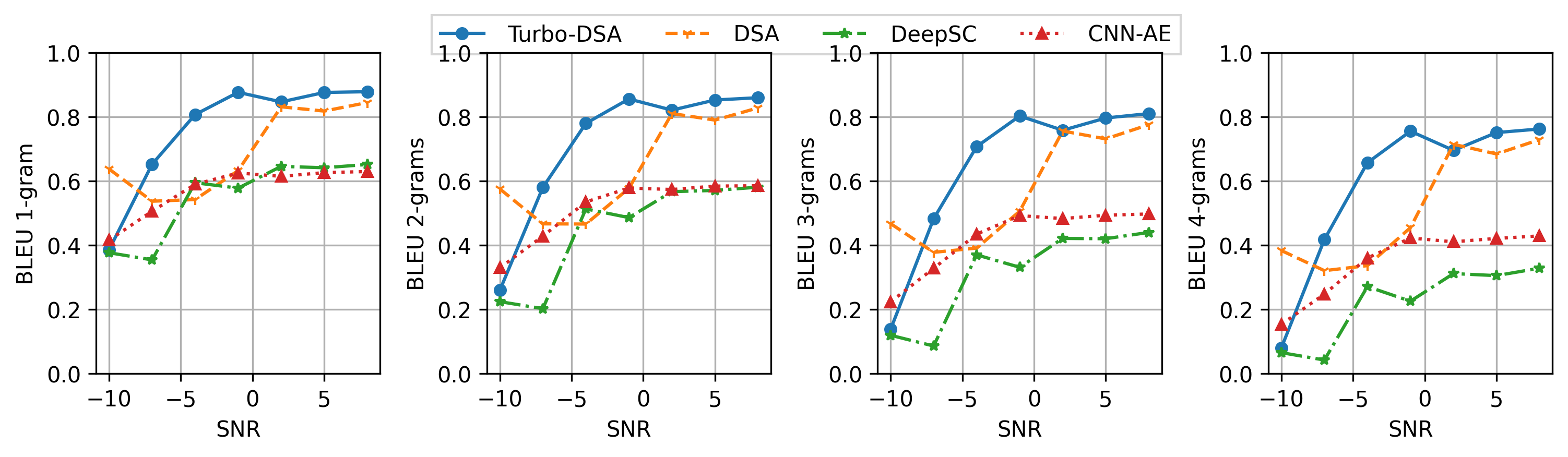}}
        
        \caption{The BLEU scores versus SNR under different Rician factors, using the maritime dataset 2.} 
        \label{147BLEU}
\end{figure*}
    
\begin{figure*}
        \centering
        \subfloat[AWGN]{\label{Simu_DataIMO_newnewRician_newnewRicianSS_K1}\includegraphics[width=0.3\linewidth]{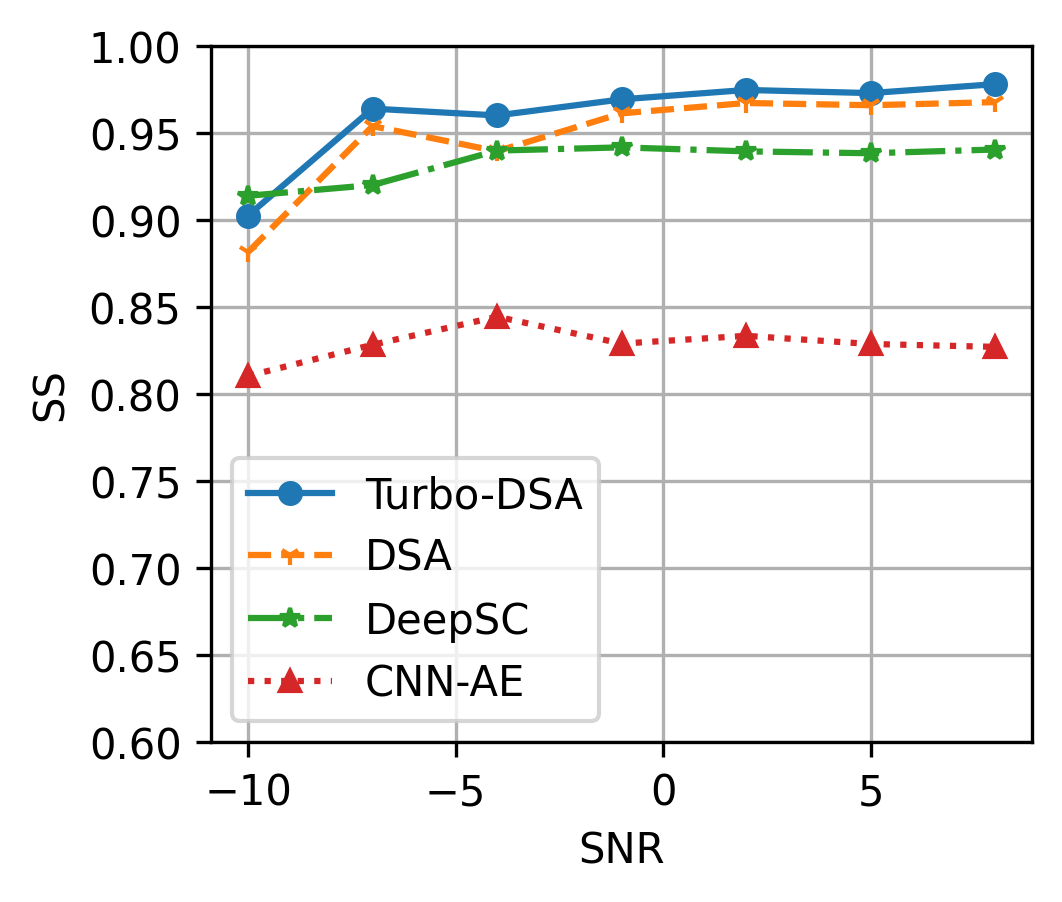}}
        \subfloat[Rician]{\label{Simu_DataIMO_newnewRician_newnewRicianSS_K4}\includegraphics[width=0.3\linewidth]{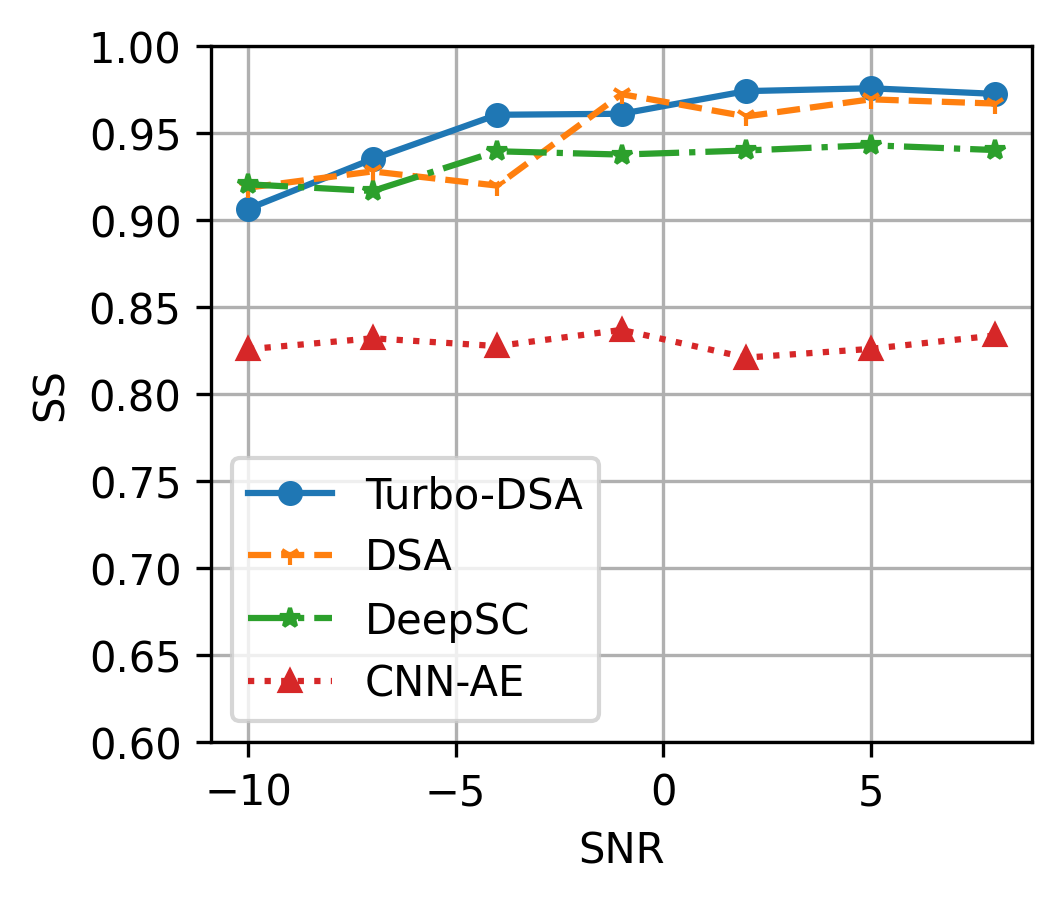}}
        \subfloat[Rayleigh]{\label{Simu_DataIMO_newnewRician_newnewRicianSS_K7}\includegraphics[width=0.3\linewidth]{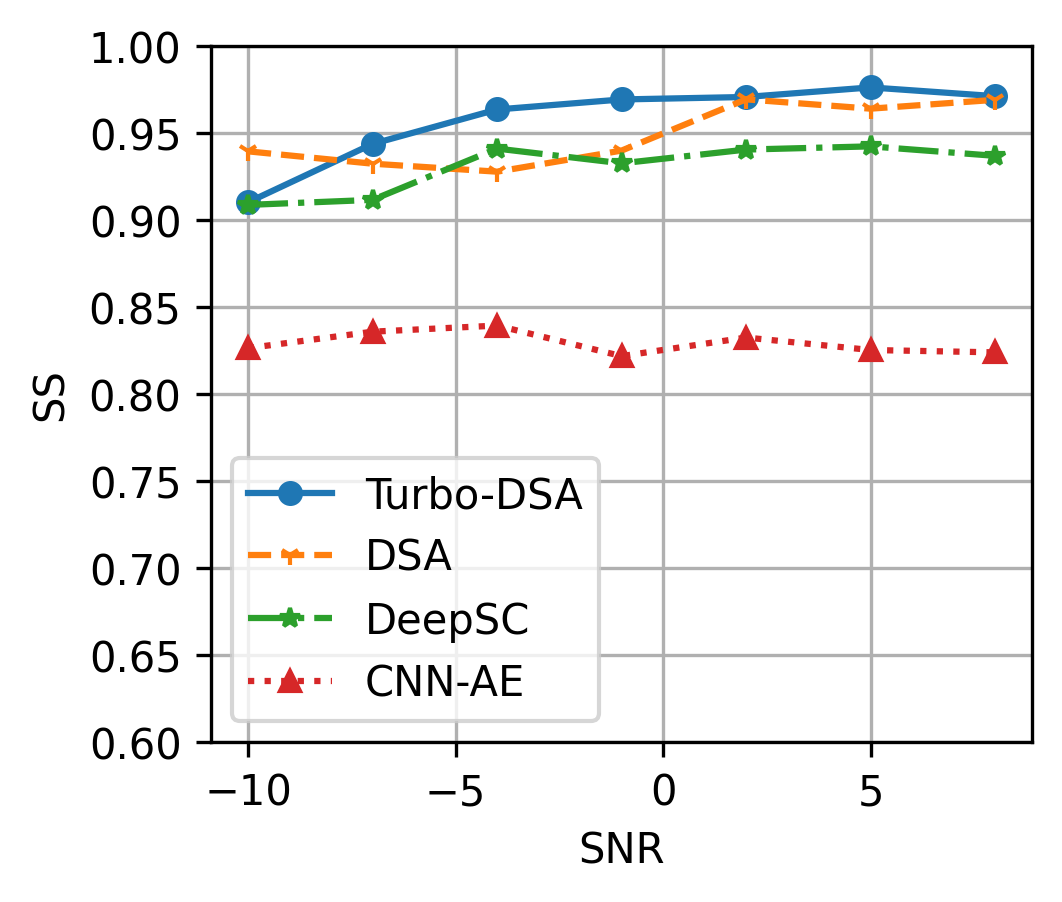}}

        \caption{The SS versus SNR under different Rician factors, using the maritime dataset 2.}
        \label{147SS}
\end{figure*}

In order to delve into the impact of Rician fading channels on the semantic transmission performance of Turbo-DSA in maritime communications, we conduct a series of experiments. Given the widespread use of Rician fading channel to simulate signal propagation characteristics in maritime environments, these experiments are crucial for assessing the effectiveness and adaptability of Turbo-DSA in real-world maritime applications.

In Fig. \ref{147BLEU}, we present the line graphs and box plots of BLEU performance for CNN-AE, DeepSC, DSA, and Turbo-DSA under various Rician factor $K$ conditions, encompassing different n-gram models (1-gram, 2-gram, 3-gram, and 4-gram), different Rician $K$ factors (1, 4, 7), and varying SNR levels.

Regarding the different n-gram models, Turbo-DSA consistently exhibits higher BLEU scores across all SNR conditions. Specifically, the 1-gram BLEU curve highlights Turbo-DSA's superior performance, achieving the highest BLEU scores across different SNR levels, particularly in low SNR conditions such as -10 dB and -7 dB. This underscores Turbo-DSA's robustness in noisy environments. Even under conditions where SNR exceeds 0 dB, Turbo-DSA maintains its lead, especially when $K$ is 4 and 7. Although CNN-AE, DeepSC, and DSA perform well under high SNR conditions, Turbo-DSA retains its advantage, demonstrating high performance at better SNRs. Moreover, Turbo-DSA maintains competitiveness in 2-gram, 3-gram, and 4-gram models, typically outperforming other systems across various SNR conditions.

In terms of different Rician $K$ factors, Turbo-DSA demonstrates strong performance stability. Particularly in low SNR conditions like -10 dB and -7 dB, Turbo-DSA achieves relatively high BLEU scores. Regardless of whether the Rician factor $K$ is 1, 4, or 7, Turbo-DSA consistently maintains high BLEU scores across all SNR conditions, highlighting its outstanding performance across various Rician factor scenarios. Additionally, Turbo-DSA's BLEU scores remain consistent across different SNR conditions, showing improved performance as SNR increases. Under high SNR conditions (8 dB), Turbo-DSA typically outperforms other systems. It retains its leading position relative to CNN-AE, DeepSC, and DSA, with even more pronounced performance advantages in low SNR conditions. This validates Turbo-DSA's exceptional performance across different noise levels, emphasizing its adaptability in noisy and challenging environments.

In Fig. \ref{147SS}, we display the Structural Similarity (SS) values of CNN-AE, DeepSC, DSA, and Turbo-DSA under different Rician $K$ factors (1, 4, 7).

For $K=1$, CNN-AE's SS values range from 0.80 to 0.85, DeepSC's SS values range from 0.90 to 0.95, DSA's SS values range from 0.88 to 0.96, whereas Turbo-DSA's SS values range from 0.90 to 0.98. Clearly, Turbo-DSA stands out in performance, with significantly higher SS values compared to other communication systems. For $K=4$ or $K=7$, Turbo-DSA's SS values range from 0.90 to 0.98. In this scenario, Turbo-DSA continues to demonstrate exceptional performance, maintaining relatively stable SS values that surpass those of other communication systems. Regardless of the $K$ factor, Turbo-DSA consistently exhibits superior performance, with SS values notably higher than other communication systems. These results clearly indicate that Turbo-DSA performs exceptionally well under different $K$ factor conditions. When the $K$ factor is 1, Turbo-DSA's SS values significantly outperform other communication systems, illustrating its excellent performance. This advantage is equally evident when the $K$ factor is 4 and 7, reconfirming Turbo-DSA's outstanding performance under various conditions.

As SNR increases, CNN-AE, DeepSC, DSA, and Turbo-DSA exhibit an upward trend in SS values, indicating enhanced capabilities in semantic extraction and recovery. Compared to CNN-AE, Turbo-DSA consistently displays superior SS performance across all SNR conditions. This is attributed to the excellence of Turbo-DSA's semantic encoder and decoder. In comparison to DeepSC and DSA, Turbo-DSA also outperforms them in SS performance. This superiority arises because Turbo-DSA's channel encoder and decoder are not simplistic neural network designs but integrate traditional Turbo encoders and decoders from communication systems, enhancing its performance. Although at SNR greater than 0 dB, CNN-AE, DeepSC, DSA, and Turbo-DSA show similar performance, at SNR less than 0 dB, Turbo-DSA exhibits superior SS performance, showcasing remarkable semantic extraction and recovery capabilities and demonstrating robustness and adaptability to extreme maritime conditions.

\subsubsection{Analysis of Network Architecture and Training Parameters}

To assess the specific impact of different network architectures and training parameters on the semantic transmission performance of Turbo-DSA, we conduct extensive simulation experiments. These experiments primarily examined how transformer layer numbers, training SNR, and learning rate affect the performance of Turbo-DSA.

\begin{figure*}
   \centering 
   \vspace*{0pt} 
   \includegraphics[width=0.9\linewidth]{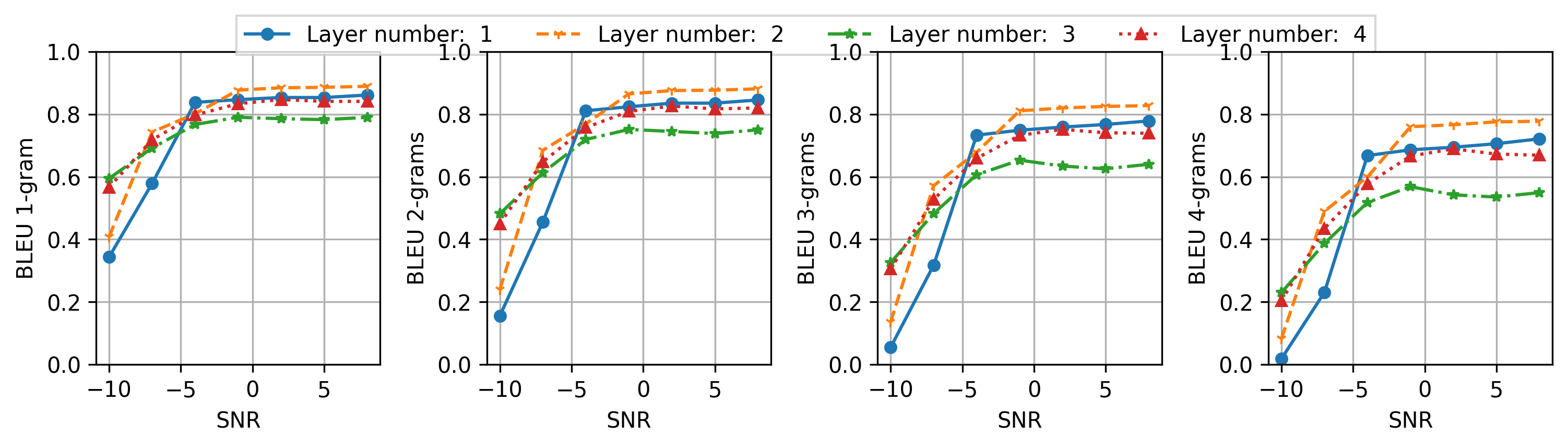}
   \caption{The BLEU scores versus SNR with Turbo-DSA under different transformer layer numbers.}
   \label{Simu_Turbo_Rayleigh_BLEU_Transformerlayer} 
\end{figure*}

Fig. \ref{Simu_Turbo_Rayleigh_BLEU_Transformerlayer} delves into the variation of BLEU scores with the increase in SNR for the Turbo-DSA system, under different network architecture configurations employed in the semantic encoder and decoder. Benefiting from the powerful capabilities of the transformer architecture, Turbo-DSA demonstrates high performance regardless of whether the number of transformer layers is set to 1 through 4. However, there remains a discrepancy of approximately 0.2 in BLEU scores between different layer counts, emphasizing the critical importance of meticulous network architecture tuning for the precise transmission of maritime text semantics in practical applications.

The simulation experiments do not simply suggest that more layers equate to better semantic transmission, nor does it imply fewer layers are inherently advantageous. Instead, the results reveal that the system achieves optimal performance with 2 transformer layers, capable of transmitting the semantics of maritime texts with relative accuracy and without error. With 4 layers, the performance is slightly inferior yet remains at a high level, possibly due to the deeper network structure facilitating the capture of more intricate semantic information. Interestingly, a 3-layer configuration does not meet expectations, potentially because an increased network depth leads to overfitting or escalates the difficulty of training, thereby impacting the accuracy of semantic transmission. Therefore, in the design and optimization of Turbo-DSA system, careful consideration of network complexity is essential, necessitating fine-tuning tailored to specific application scenarios to ensure optimal performance in the transmission of maritime text semantics.

\begin{figure*}
   \centering 
   \vspace*{0pt} 
   \includegraphics[width=0.9\linewidth]{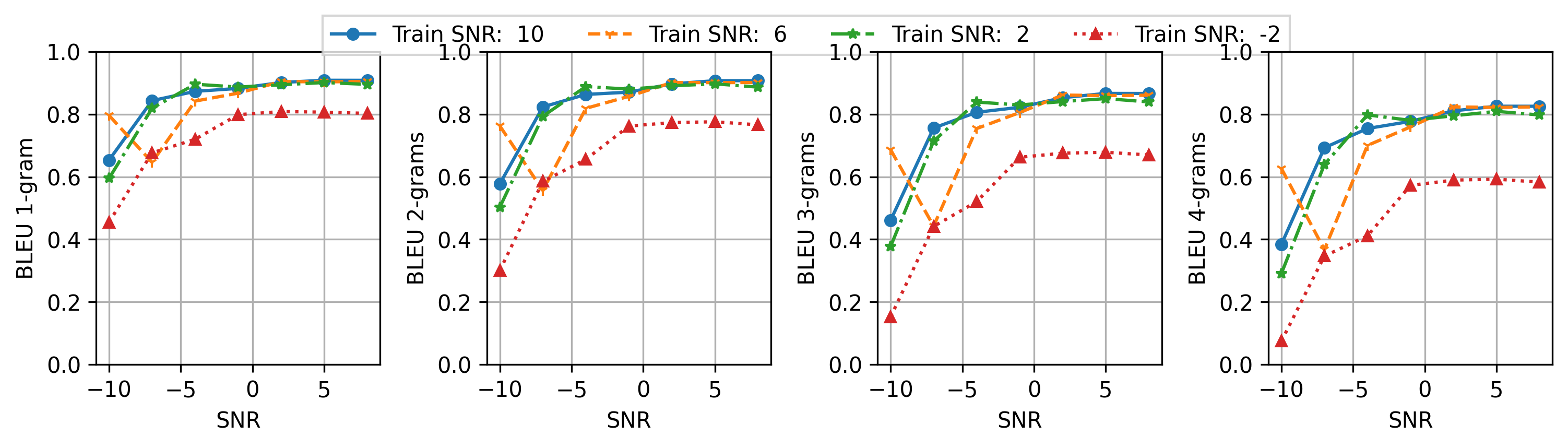}
   \caption{The BLEU scores versus SNR with Turbo-DSA under different train SNR conditions.}
   \label{Simu_Turbo_Rayleigh_BLEU_TrainSNR} 
\end{figure*}

Fig. \ref{Simu_Turbo_Rayleigh_BLEU_TrainSNR} reveals the trend of BLEU scores for the Turbo-DSA system varying with the test SNR under different training SNR conditions. The observations highlight that the disparity in training SNR significantly impacts BLEU scores, underscoring the importance of training SNR in enhancing system performance. Given that Turbo-DSA is aimed at addressing the challenges of marine IoT communications, and due to the complex nature of maritime channels, the training SNRs selected in this work are generally lower than those in other related works.

Taking the 1-gram evaluation of BLEU as an example, the Turbo-DSA system demonstrates superior performance when trained at SNRs of 2 dB, 6 dB, and 10 dB. Particularly noteworthy is the system's ability to maintain high BLEU scores even when the test SNR is below 0 dB, showcasing its robustness in adverse communication environments. More crucially, even with a training SNR as low as -2 dB, the BLEU performance of Turbo-DSA remains outstanding, exhibiting a clear advantage over the comparative systems shown in Figure 7. These findings indicate that Turbo-DSA possesses unique strengths in terms of adaptability and effectiveness to training SNR, making it particularly suitable for handling semantic communication of maritime texts.

\begin{figure}
   \centering 
   \vspace*{0pt} 
   \includegraphics[width=0.5\linewidth]{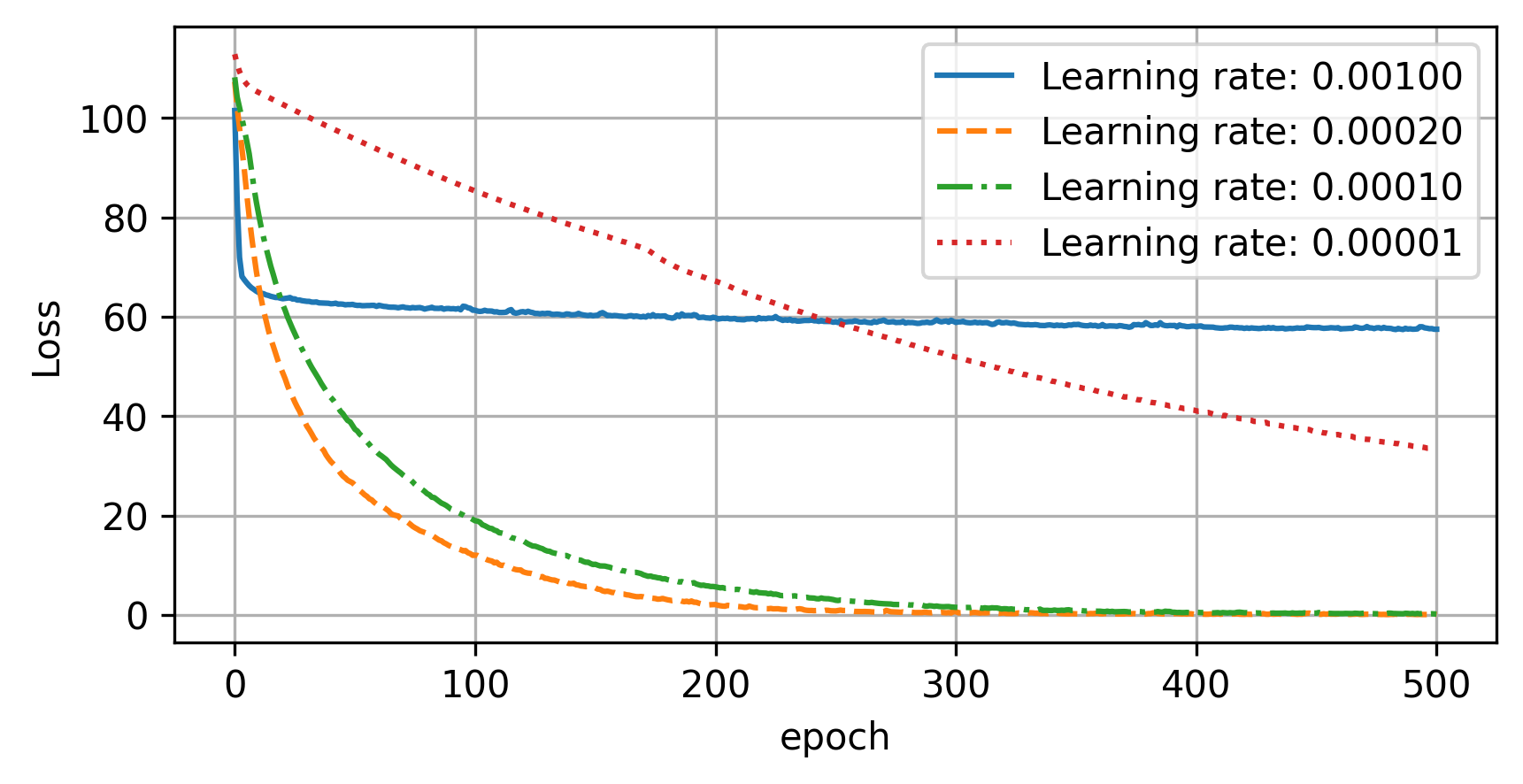}
   \caption{The loss values versus SNR with Turbo-DSA under different train learning rates.}
   \label{Simu_loss_Rayleigh_lr} 
\end{figure}

\begin{figure*}
   \centering 
   \vspace*{0pt} 
   \includegraphics[width=0.9\linewidth]{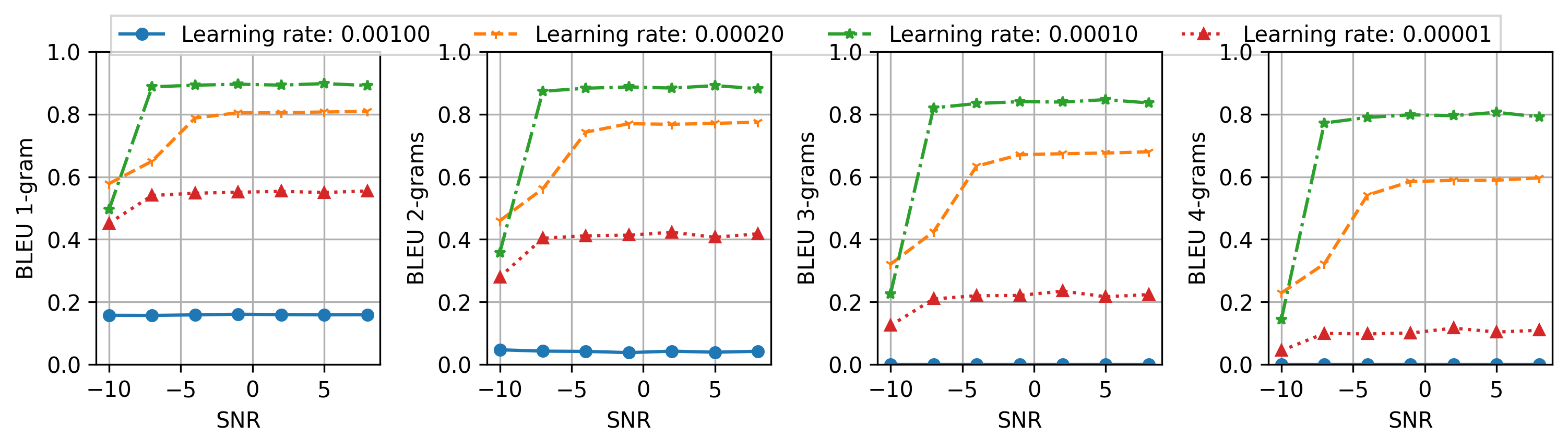}
   \caption{The BLEU scores versus SNR with Turbo-DSA under different train learning rates.}
   \label{Simu_Turbo_Rayleigh_BLEU_lr} 
\end{figure*}

Fig. \ref{Simu_loss_Rayleigh_lr} illustrates how the training loss values of Turbo-DSA vary with the increase in epochs under different learning rates. Concurrently, Fig. \ref{Simu_Turbo_Rayleigh_BLEU_lr} documents the trend of BLEU scores changing with the increase in SNR under identical settings. Both figures are based on models with the same network architecture but trained with varying learning rates.

It is evident that different learning rates lead to significant differences in the loss function values, highlighting the decisive impact of learning rate on the convergence speed and ultimate performance of the model. Specifically, when the learning rate is set at 0.00100, the model fails to learn effectively, as indicated by the failure of the loss value to drop sufficiently. Conversely, when the learning rate is reduced to 0.00001, the model may suffer from overfitting, making it extremely difficult for the loss value to converge. In contrast, setting the learning rate to 0.00010 or 0.00020 allows the model to learn more efficiently, with the loss value gradually stabilizing, demonstrating good convergence properties. This indicates that the model has successfully captured the core semantic features of the training data.

From Fig. \ref{Simu_Turbo_Rayleigh_BLEU_lr}, it becomes clear that BLEU scores are closely tied to the selection of learning rate. When the learning rate is 0.00100, the BLEU score is low, indicating that the model is unable to accurately convey the semantic information of maritime texts. Similarly, when the learning rate is too low, at 0.00001, the model still fails to transmit semantics successfully. The consistent results across both figures underscore the critical importance of an appropriate learning rate for the model to learn maritime text semantics. Especially at a learning rate of 0.00010, the model demonstrates optimal learning performance, able to convey semantic information with the highest accuracy. Although a learning rate of 0.00020 also yields satisfactory results, its BLEU score falls short of that achieved at 0.00010. This suggests that even when a neural network converges to a stable state, the gradient might still get stuck in a local minimum rather than reaching the global minimum, affecting the model's final performance.

In summary, Turbo-DSA demonstrates superior BLEU and SS performance compared to the benchmarks. Its outstanding performance can be attributed to the use of semantic encoder and decoder designed by transformer, which sets it apart from CNN-AE, showcasing exceptional semantic extraction and reconstruction capabilities. In comparison to DeepSC and DSA, Turbo-DSA's channel encoder and decoder do not solely rely on pure neural network concatenation but draw inspiration from the structure of Turbo encoders and decoders, enabling superior semantic extraction and recovery of maritime data, especially under low SNR conditions. Furthermore, owing to its unique structure, Turbo-DSA exhibits greater robustness under different Rician factor $K$ conditions, making it more suitable for challenging maritime environments.

\section{Conclusion}
\label{Conclusion}

In order to achieve even higher efficiency in communication networks, this paper has introduced a novel Turbo-DSA scheme, which integrates the Turbo principle into the semantic encoding and decoding processes, providing robust performance for semantic communication among diverse marine IoT devices. Turbo-DSA employs transformers to construct semantic vectors, and utilizes DL-based Turbo framework to further protect these vectors. Extensive simulation results have demonstrated the superior performance of Turbo-DSA across various metrics. Notably, under low SNR conditions, Turbo-DSA exhibits significant performance advantages. Beyond efficient semantic extraction and recovery in maritime communications, Turbo-DSA shows exceptional adaptability to the demanding marine environment. However, the component encoding and iterative decoding processes in Turbo-DSA require a significant amount of computation, which may limit the application of the algorithm in resource-constrained environments. In the future, we will consider semantic transmission in marine IoT with multiple nodes, aiming to reduce computational load and enhance the overall network efficiency.


\section*{Acknowledgement}

The work was supported by the National Natural Science Foundation of China (No. 51939001, No. 62371085) and Fundamental Research Funds for the Central Universities (No. 3132023514).
\bibliographystyle{cas-model2-names}
\bibliography{Ref-TurboDSA}

\end{document}